\documentclass[twocolumn]{aastex631}
\usepackage{bm}
\usepackage{amsmath}
\usepackage{soul}
\def\la{\hbox{\raise.35ex\rlap{$<$}\lower.6ex\hbox{$\sim$}\ }}
\def\ga{\hbox{\raise.35ex\rlap{$>$}\lower.6ex\hbox{$\sim$}\ }}

\def\zunit{\hat {\bf z}}
\def\yunit{\hat {\bf y}}
\def\xunit{\hat {\bf x}}

\def\St{{\rm {St}}}

\def\setV{{\cal X}^{^{\{v\}}}}
\def\setVN{{\cal X}^{^{\{vn\}}}}
\def\setF{{\cal X}^{^{\{f\}}}}
\def\setFN{{\cal X}^{^{\{fn\}}}}
\def\beq{\begin{equation}}
\def\eeq{\end{equation}}
\def\beqa{\begin{eqnarray}}
\def\eeqa{\end{eqnarray}}

\def\order#1{{\cal O}\left({#1}\right)}

\newcommand{\NewEditv}[1]{{{\color{black} #1}}}

\newcommand{\REV}[1]{#1}

\definecolor{darkgreen}{HTML}{25911E}

\definecolor{purple}{rgb}{0.7,0.0,0.7}

\newcommand{\mstar}{M_{\star}}
\newcommand{\alphat}{\Tilde{\alpha}}

\received{\today}
\submitjournal{Astrophysical Journal}

\shorttitle{Evidence for turbulent concentration}
\shortauthors{Umurhan et al.}
\graphicspath{{./}{}}

\begin{document}

\title{Evidence For Turbulent Concentration In Particle-Laden Midplane Layers of Planet-Forming Disks.}

\author[0000-0001-5372-4254]{Orkan M. Umurhan}
\affiliation{SETI Institute, 389 Bernardo Way, Mountain View, CA 94043, U.S.A.}
\affiliation{Space Sciences Division, Planetary Systems Branch, NASA Ames Research Center,  Mail Stop 245-3, Moffett Field, CA 94035, USA}
\correspondingauthor{Orkan Umurhan}
\email{oumurhan@seti.org}

\author[0000-0003-0801-3159]{Debanjan Sengupta} 
\affiliation{Department of Astronomy, New Mexico State University, Las Cruces, NM 88003-8001, USA}

\author[0000-0002-5062-9507]{Thomas Hartlep} 
\affiliation{Bay Area Environmental Research Institute, Moffett Field, CA 94035, USA}

\author[0000-0003-0553-1436]{Jeffrey N. Cuzzi} 
\affiliation{SETI Institute, 389 Bernardo Way, Mountain View, CA 94043, U.S.A.}
\affiliation{Space Sciences Division, Planetary Systems Branch, NASA Ames Research Center,  Mail Stop 245-3, Moffett Field, CA 94035, USA}
\author[0000-0002-1132-5594]{Paul R. Estrada} 
\affiliation{Space Sciences Division, Planetary Systems Branch, NASA Ames Research Center,  Mail Stop 245-3, Moffett Field, CA 94035, USA}


\defcitealias{Umurhan_etal_2020}{U+20}
\defcitealias{Yang_etal_2017}{Y+17}
\defcitealias{Carrera_etal_2015}{C+15}
\defcitealias{Li_Youdin_2021}{LY21}
\defcitealias{Estrada22_paper2}{E+22}
\defcitealias{Schafer_Johansen_2022}{SJ22}
\defcitealias{Sengupta_Umurhan_2023}{SU23}
\defcitealias{Squire_Hopkins_2018a}{SH18a}
\defcitealias{Sengupta_etal_2024}{S+24}
\defcitealias{Hartlep_etal_2017}{H17}
 
\begin{abstract}





\REV{We investigate the axisymmetric, weakly turbulent state of settled particle layers in a model of a globally laminar protoplanetary disk. We focus on conditions in which the large-scale axisymmetric filaments associated with the streaming instability (SI) either cannot form or have not developed yet. We observe small-scale particle clumping consistent with turbulent concentration (TC), in which
particle-rich filaments align with regions of high gas strain rate and enclose gas-only voids exhibiting coherent vorticity. Across a range of
particle Stokes numbers, $\St_K$ ($0.01-0.04$)—defined as stopping times relative to the Keplerian frequency—effective Stokes number within particle voids, $\St_\omega$, defined instead using local gas vorticity, collapses to values in the range
$\sim 0.4-0.7$ for the $\St_K$ considered. These values lie close to critical turbulent Stokes numbers associated with maximal clustering intermittency identified in statistical studies of TC. A timescale comparison reveals that in simulations with midplane particle-to-gas density ratios below unity and $\St_K \ll 1$, SI growth rates are 1–2 orders of magnitude slower than turbulent overturn frequencies at the large-eddy scale, which appears to
rule out SI as the primary driver of turbulence here. Instead, we suggest the Symmetric Instability (SymI) may be responsible. We show for our St$_K$ that
TC is a persistent feature of our turbulent particle layers, and conclude that small scale particle density fluctuations exceeding the Roche density {\it within} large-scale axisymmetric SI filaments reported in the literature are
also expressions of TC operating on top of
the slightly elevated background particle densities within those large-scale structures.}

\end{abstract}

\keywords{Protoplanetary disks, Multiphase Turbulence, Planetesimal Formation, Turbulent Concentration}

\section{Introduction} \label{sec:intro}

 When inertial particles are present in a turbulent carrier fluid, the particles tend to cluster in regions with low vorticity and high strain rate in a size-dependent way \citep{Squires_Eaton_1991, Balachandar_Eaton_2010, Bragg_Collins_2014, Bec_etal_2023} leading to strong inhomogeneity in the particle density across flow scales. This robust phenomenon is known as Turbulent Concentration (TC hereafter) and is involved in observations of many varied physical systems (rain initiation, reactive dispersed flow including combustion chambers,  etc.) and also in numerical studies of pure fluid flows exhibiting isotropic homogeneous turbulence \citep{Squires_Eaton_1990, Squires_Eaton_1991, Elghobasi_Truesdell_1992, Yang_Lei_1998, Wang_Squires_1996}.  TC has been numerically modeled to explain certain properties of tiny grains in turbulent molecular clouds \citep{Hopkins_Lee_2016} and examined in detail through statistical analyses informed by fluid dynamics simulations \citep{Hartlep_etal_2017}. In the context of planetesimal formation, \NewEditv{TC has also been explored using statistical models to extrapolate numerical simulation results to the relevant scales \citep{Cuzzi_etal_2001,Cuzzi_etal_2008,Cuzzi_etal_2010,Pan_etal_2011,Hopkins_2016a,Hopkins_2016b,Gerosa_etal_2023}. The work presented here, however, is the first to demonstrate TC directly under nebula conditions \REV{(even if globally laminar)} at realistic length scales and with sufficient numerical resolution to capture the essential dynamics.}

 \REV{In protoplanetary disks}, solid particles settle towards the midplane under the influence of the vertical component of the stellar gravity and generate a particle-gas shear layer leading to a \REV{locally} turbulent state, as originally envisioned by \citet{Weidenschilling_1980}. Later on, \citet{Cuzzi_etal_1993} and \citet{Dobrovolskis_etal_1999} predicted the generation of an Ekman-type flow structure in these shearing particle layers owing to the strong rotation present in the disk. Recently, \citet[][hereafter \citetalias{Sengupta_Umurhan_2023}]{Sengupta_Umurhan_2023} confirmed these predictions through moderate to high resolution particle-gas simulations in which they demonstrate that a combination of three separate instabilities are responsible for the turbulence generated by the particle laden midplane layer: 1) the Kelvin-Helmholtz instability (KHI) arising from the vertical variation of radial flow field; 2) the Symmetric Instability \citep[SymI;][under the single fluid approximation]{stamper_Taylor_2017, Sengupta_Umurhan_2023} arising from the vertical variation of the azimuthal flow field; and finally 3) the Streaming Instability \citep[SI;][]{Youdin_Goodman_2005} rooted in the differential drift of particle and gas fields, present in the form of axisymmetric filamentary bands of solid overdensities, when the parameters are favorable. Of these, SI has been the favored mechanism for the formation of the self-gravitating bound clumps that \REV{exceed Roche density and lead to} 
 planetesimals \citep[][among many others]{Johansen_etal_2007, Yang_etal_2017} under conditions where the nebula is marginally laminar and/or particles are bigger than implied by observations \citep{Carrascogonzalez_etal_2019, Tazzari_etal_2021}, growth models \citep{Estrada_etal_2016, Estrada_etal_2022,Sengupta_etal_2019}, and meteoritic evidence \citep[][and references therein]{Simon_etal_2018}. 

The SI is both subtle and manifold in its mechanical expression. As a resonant momentum exchange between the particle and gas fields \citep{Squire_Hopkins_2018a}, it operates across the full spectrum of vertical wavenumbers, though it is limited to radial scales above a certain threshold. The SI's fastest linear growth occurs \REV{for particle- to gas-density ratios exceeding one and} at the shortest vertical wavelengths with rates on the order of the local disk rotation frequency \citep{Youdin_Goodman_2005,Youdin_Johansen_2007}. In globally laminar disk models—such as those considered here—these ultra-short vertical wavelength modes dominate the generation of turbulence \REV{in unstratified settings}, as shown in the 
analysis of \citet{Baronett_etal_2024}. Longer vertical wavelengths, while also unstable, grow 
more slowly and 
give rise to broader radial-scale azimuthally aligned filaments. These filaments are the sites of high-amplitude particle density fluctuations, as has been reported throughout the literature. However, their development is sensitive to the background turbulence level. If the turbulence is too strong, the filaments can be weakened, lose spatial coherence, or become suppressed entirely \citep[e.g.,][]{Umurhan_etal_2020,Chen_Lin_2020}.

 In situations where turbulence is generated self-consistently in, \NewEditv{and locally around} the particle-laden midplane layer of the protoplanetary disk (irrespective of mechanism), the two necessary conditions for TC -- the presence of turbulence and inertial particles therein -- are automatically satisfied, and there is no reason to think that this mechanism refrains from taking part in the process of particle clumping. Recently, \citet{Hartlep_Cuzzi_2020} presented a statistical study based on \REV{high-resolution} homogeneous and isotropic 3D turbulence \REV{simulations} 
 \citep{Bec_etal_2010} showing that the realization of the maximum clumping by TC (maximum intermittency in the particle density field) requires certain scales to be resolved. However, even in lower-resolution numerical simulations, TC nevertheless still operates and contributes to particle clumping to a finite extent, although not attaining the maximum level of clustering. In simulations like these, we expect that there is always some contribution from TC to particle clumping that is often entirely attributed to the SI. Some clarity on this matter is needed.

In this spirit, we analyze 3D axisymmetric simulations from \citet{Sengupta_Umurhan_2023}, conducted under \REV{globally} 
laminar conditions ({\it i.e.}, without any external sources of turbulence) and show that the self-generated turbulence in the particle-\REV{rich} 
midplane layer exhibits TC \NewEditv{where its fingerprints can be seen in earlier published models \citep[e.g.][]{Yang_etal_2017}.} We focus on parameter regimes where the 
turbulence is strong enough to suppress the emergence of any azimuthally aligned SI filaments, 
\REV{such as} identified in previous work as \NewEditv{the earliest manifestations of SI, and} the sites of high-density fluctuations capable of exceeding the Roche threshold. We examine the statistics and structure of this turbulent configuration \REV{to identify} 
how turbulent concentration of particles manifests within the flow.

\section{Equations of motion, Solution Method, numerical experiments conducted and input parameters}\label{sec:EQNSandmore}

\begin{deluxetable}{cl}
\label{tbl:definitions}
\tabletypesize{\footnotesize}
\tablecolumns{2}
\tablewidth{0pt}
\tablecaption{Variables used in theoretical modeling}
\tablehead{
\colhead{Variable} & \colhead{Definition}}
\vspace{-0.2cm}
\startdata
\vspace{-0.2cm}\\
$\bm{U}$ & Total gas velocity vector\\
$\bm{V}$ & Total particle velocity vector\\
$\bm{U}_{cm}$ & Center of mass velocity in lab frame\\
$U_{{\rm rms}}$ & rms gas speed \\
$U_K$& Keplerian speed \\
$c_s$& Local sound speed \\
$\eta$& Normalized radial pressure gradient \\
$\Pi$& Dynamical compressibility \\
$\Omega_K$ & Keplerian frequency \\
$\Omega_L$ & Large eddy frequency \\
$\Omega_{\ell}$ & Frequency of eddy with length-scale $\ell$ \\
$R'$ & $\Omega_L / \Omega_K$  \\
${\rm Ri}_{\phi}$ & Richardson number for azimuthal flow \\
$\delta V_0$ & Amplitude of the gas azimuthal velocity shear \\ 
$H$ & Gas scale height ($\equiv c_s/\Omega_K$)\\
$h$& Particle scale height\\
$H_s$ & FWHM of the azimuthal shear\\
$P$ & Gas pressure\\
$L_x,L_z$& Simulation box size (in $H$) in $x$ \& $z$ \\
$k_L$ & Wavenumber for driving scale \\
$k_h$ & Wavenumber for particle scale height $h$ \\
$k_{\mu\epsilon}$& Radial wavenumber of fastest growing mode \\
$\alphat$ & Turbulent intensity: $U_{{\rm rms}}^2/c_s^2$ assuming $\Omega_K=\Omega_L$\\
$\alpha$ & Turbulent viscosity from mixing length theory: $\alphat / R'$ \\
$\alpha_h$ & Turbulent viscosity derived from particle scale height \\
$\rho_g~(\rho_p)$ & Gas (particle) density\\
$\overline\rho_p $ & Radial average of $\rho_p$ \\
$\epsilon$ & Local dust-to-gas mass ratio $(\rho_p/\rho_g)$\\
$\epsilon_{_0}$ & $\epsilon$ at midplane, $z=0$. \\
$t_s$ & Aerodynamic stopping (friction) time for particles \\
${\rm St}_K$ & Stokes number \textit{wrt} Keplerian $(t_s\Omega_K)$\\
${\rm St}_L$ & Stokes number \textit{wrt} large eddy $(t_s\Omega_L)$\\
${\rm St}_{\ell}$ & Stokes number \textit{wrt} eddy of length $\ell$ $(t_s\Omega_{\ell})$\\
$R$ & Orbital distance from central star \\
$S_{ij}, S$ & Strain rate tensor and magnitude \\
$\bm{\omega}, \omega$ & Gas vorticity and magnitude \\
$\varepsilon$ & Gas spectral kinetic energy per unit gas mass \\
$\setV$ & Set of grid pts. where $\rho_p = 0$ and $|z|<z_{{\rm lim}}$ \\
$\setF$ & Set of grid pts. where $\rho_p \neq 0$ and $|z|<z_{{\rm lim}}$ \\
$\setVN$ & Like $\setV$ with all neighboring pts. also empty \\
$\setFN$ & Like $\setF$ with all neighboring pts. contain particles  \\
$z_{\rm{lim}}$& Scale height of void-filament complex confinement \\
$Z$& Disk metallicity \\
\vspace{-0.2cm}\\
\enddata
\end{deluxetable}

The coupled particle-gas flow field is governed by the mass and momentum conservation equations, which in the cylindrical coordinate system $(\hat{R},\hat{\phi},\hat{z})$ with unit vector $\hat{\bf r}=R\hat{\bf R}+\phi\hat{\bf{\phi}}+z\hat{\bf z}$, read as:
\begin{subequations}
\begin{equation}\label{eqn:gascontinuity}
\frac{\partial \rho_g}{\partial t}+\nabla\cdot\left(\rho_g \bm{U} \right)=0;
\end{equation}
\begin{equation}\label{eqn:gasmomentum}
\frac{\partial \bm{U}}{\partial t} +\left(\bm{U}\cdot\nabla\right)\bm{U} = -\Omega_K^2{\bf \hat{r}} +\frac{\rho_p}{\rho_g}\frac{\bm{V} - \bm{U}}{t_s}-\frac{1}{\rho_g}\nabla P,
\end{equation}
\end{subequations}
where $\bm{U} \equiv (u_g, v_g, w_g)$ is the three component gas velocity 
and $\bm{V}$ is the  velocity of the particle field. The gas and particle volume densities are $\rho_g$ and $\rho_p$, $P$ is the gas pressure, $\Omega_K=\sqrt{G\mstar/R^3}$ where $G$ and $\mstar$ are the universal gravitational constant and the mass of the central star, respectively. The Keplerian velocity is then expressed as $U_K = R\Omega_K$.
Treating particles as the second fluid, the same equations are written as 
\begin{subequations}
\begin{equation}\label{eqn:dustcontinuity}
\frac{\partial \rho_p}{\partial t}+\nabla\cdot \left(\rho_p\bm{V}\right)=0;
\end{equation}
\begin{equation}\label{eqn:dustmomentum}
\frac{\partial\bm{V}}{\partial t}+\left(\bm{V}\cdot \nabla\right)\bm{V}=-\Omega_K^2 {\bf \hat{r}} -\frac{\bm{V} -\bm{U}}{t_s}.
\end{equation}
\end{subequations}
\noindent
The second term on the RHS of Eqs. (\ref{eqn:gasmomentum}) and (\ref{eqn:dustmomentum}), being proportional to the relative velocities of the gas and the particle fields, denotes the aerodynamic drag, with $t_s$ being the 
particle stopping time. 
The particle flow field is considered pressure-less and moves with the local Keplerian velocity \NewEditv{in the absence of gas}. 
Gas, on the other hand, feels the radial pressure gradient ($\nabla P$ term in Eq. (\ref{eqn:gasmomentum})), which makes its motion slightly sub-Keplerian. In steady state we assume that the background pressure field has a radial gradient given by $\nabla P = -\rho_g \eta R$, where
\begin{equation}\label{eqn:eta}
\eta=-\frac{1}{2}\left(\frac{H}{R}\right)^2\frac{\partial \ln\rho_g}{\partial \ln R}=\frac{1}{2}\beta \left(\frac{H}{R}\right) > 0,
\end{equation}
\NewEditv{is the normalized pressure gradient}, $H/R$ is the disk aspect ratio 
in terms of the gas scale height $H=c_s/\Omega_K$, and the parameter $\beta$ is 
\begin{equation}\label{eqn:beta}
\beta=-\left(\frac{H}{R}\right)\frac{\partial \ln \rho_g}{\partial \ln R}.
\end{equation}
The dynamical compressibility of the system is represented by the parameter $\Pi=\eta U_K/c_s$. With Eqs. (\ref{eqn:eta}-\ref{eqn:beta}), $\Pi$ can be expressed as 
\begin{equation}
    \Pi=\frac{1}{2}\beta = \frac{R}{H}\eta.
\end{equation}
\NewEditv{For all simulations performed we have adopted $\Pi = 0.05$.}

\begin{deluxetable*}{cccccccc}
\label{tbl:simulation}
\tablecolumns{7} 
\tablecaption{List of simulations and relevant parameters.}
\tablehead{ 
\colhead{Identifier} & \colhead{$L_x, L_y, L_z ^{a}$} & \colhead{$N_x$, $N_y$, $N_z$} & \colhead{${\rm St}_K$ } & \colhead{$Z$} & $\nu_{\rm p}$ & $\rho_{\rm par,swarm}\big/\rho_g\ ^{b}$ & $t\Omega_k ^{b,c}$ 
}
\startdata 
\vspace{-0.2cm}\\
St-01.Z-02 & $0.4 \times 0.4 \times 0.4$  & $4096 \times 1 \times 4096$ & $0.01$ & $0.02$ & $1/4$ & 0.500 & 400 \\
St-02.Z-01 & $0.4 \times 0.4 \times 0.4$  & $4096 \times 1 \times 4096$ & $0.02$ & $0.01$ & 1/4 & 0.250 & 300 \\
St-04.Z-01& $0.2 \times 0.2 \times 0.2$ & $2048 \times 1 \times 2048$ & $0.04$ & $0.01$ & 1 & 0.125  & 250  \\
\vspace{-0.2cm}\\
\hline
\vspace{-0.2cm}\\
\enddata 
\tablenotetext{a}{ In units of $H$.}
\tablenotetext{b}{ Values are rounded.}
\tablenotetext{c}{ Corresponds to timestamp of the snapshot analyzed.}
\end{deluxetable*}

We solve Eqs. (\ref{eqn:gascontinuity}-\ref{eqn:dustmomentum}) using the {\sc Pencil Code}\footnote{\url {http://pencil-code.nordita.org/ }} \citep{Pencil_code},  a higher order (sixth order in space and third order in time) finite difference code in a shearing box setup, which considers dynamics in a box centered around the reference radial position at $R$ with the shear term being linearized. For details of the shearing box setup of the fluid equations above, the reader is directed to Sec. 2 of \citetalias{Sengupta_Umurhan_2023}.

Our simulations are 3D-axisymmetric, where the gradients of the wave vectors in the azimuthal direction are not considered. We use a sixth-order hyper-viscosity and hyper-diffusion in order to allow for the fields to dissipate their energy near the smallest scale while preserving the power spectra at the large scales.

For simulations with mono-disperse solids, we use Lagrangian super-particles, each of which represents a swarm of identical particles with a single Stokes number $\St_K$, interacting with the gas collectively through the drag force. The properties of super-particles are set based on the parameters used for the simulation. The mean gas density in the box is $\rho_m=\Sigma_g/L_z$, where $\Sigma_g$ is the vertically integrated gas surface density of the box and $L_z$ is the vertical size of the computational domain. The representative density of each super-particle thus reads 

\begin{equation}
\rho_{\rm par,swarm}=\frac{Z\rho_m}{N_{\rm par}/(N_xN_yN_z)},
\label{swarm_definition}
\end{equation}
\noindent where $N_{par}$ is the total number of super-particles introduced in the box with number of grids $N_x$, $N_y$ and $N_z$ in the $x$, $y$ and $z$ directions, respectively. Thus, $\nu_{\rm p} = N_{{\rm par}}/(N_xN_yN_z)$ is the number of particles per grid point for a given simulation. Similarly, the total mass represented by each super-particle is given by

\begin{equation}
m_{\rm par,swarm}=\frac{Z\rho_m V_{\rm box}}{N_{\rm par}},
\end{equation}
\noindent where $V_{box}=L_xL_yL_z$ is the volume of the simulation box.  \par

For the simulation setup, we use a domain with size $0.2H$ in the radial and vertical direction with a grid size $4096 \times 4096$ for St$_K=0.01$ and $0.02$, and $2048 \times 2048$ for St$_K=0.04$. The list of simulations with the identifiers and all the relevant initial parameters are listed in Table \ref{tbl:simulation}. We implement a shearing boundary condition in the radial direction and a reflective boundary condition in the vertical direction. The initial gas density and local sound speed is set as $\rho_{g,0} = c_{s,0} = 1$ along with $\Omega_{K,0}=1$, which sets the length scale $H = 1$. \NewEditv{With these inputs and in the absence of particles $u_g = w_g = 0$ and $v_g = -0.05 c_s$ everywhere, signifying a weakly radially pressure-supported Keplerian steady state.}


 \section{Relevant definitions}\label{sec:definitions}
 In this section, we present the important physical quantities along with the methodologies and formalisms we shall adhere to for the purpose of the analysis presented in this paper. 
 
\subsection{Vorticity and strain-rate}\label{sec:def_1}
 
To start with, we define the gas vorticity \NewEditv{in physical space} as 
\beqa 
& & {\bm{\omega}} \equiv  
\nabla \times {\bm{U}} = 
- \partial_z v_g\xunit + 
\nonumber \\
& & \ \ \ \ 
\left(\partial_z u_g-\partial_x w_g\right)\yunit
+
\left(-3\Omega_K/2 + \partial_x v_g\right)\zunit,
\eeqa
where the amplitude of ${\bm{\omega}}$ is given by $\omega \equiv \sqrt{\omega_i\omega_i}$,
in which the Einstein summation convention is implied.  

We similarly define and later employ the symmetric strain rate tensor
\beq
S_{ij} = \frac{1}{2}\left(
\frac{\partial U_{j}}{\partial x_i}+
\frac{\partial U_{i}}{\partial x_j}
\right).
\eeq
There are several ways to define a scalar measure of $S_{ij}$. A common one is defined by its magnitude, i.e., 
$S \equiv \sqrt{S_{ij} S_{ji}}$.
Here, we adopt a more traditional approach and define the {\it scalar strainrate} by
\beq
S \equiv ||S_{ij}||,
\eeq
where $||S_{ij}||$ is the matrix norm, taken as the largest absolute eigenvalue of $S_{ij}$ (since $S_{ij}$ is symmetric). In practice, we compute $S_{ij}$ at each grid point $(x, z)$, evaluate its eigenvalues $\sigma_{1,2,3}$, and assign $S$ to be the one with the greatest absolute value.\par


\subsection[Particle scale-height \texorpdfstring{$h$}{h}]{Particle scale-height $h$}\label{sec:par-scale-height}

Measurement of the particle scale height $h$ from the simulation data is subtle as demonstrated in \citet[][\citetalias{Sengupta_etal_2024} hereafter]{Sengupta_etal_2024}. Generally, $h$ is measured using the relationship

\beq
h^2 = \int_{-\infty}^{\infty}z^2\overline\rho_p(z) dz\bigg/\int_{-\infty}^{\infty}\overline\rho_p(z) dz,
\eeq
where $\overline\rho_p(z)$ is the radial mean of particle density defined as
\beq  
\overline\rho_p(z) \equiv \frac{1}{L_x} \int_{-L_x/2}^{L_x/2}\rho_p(x,z) dx.
\label{eq:radial_average_rhop}
\eeq
In practice,
\beq
h^2 \approx \sum_{i=1}^{N_z} z_i^2 \overline\rho_p(z_i)\Big/ \sum_{i=1}^{N_z} \overline\rho_p(z_i),
\eeq
in which $z_i$ is the vertical grid point labeled by $i$. However, \citetalias{Sengupta_etal_2024} showed that this method generally overestimates the value of $h$, especially when there are particles present near the vertical boundary, a situation particularly true for small Stokes numbers as in the simulations here. 
\NewEditv{Under such conditions, \citetalias{Sengupta_etal_2024} showed that the most reliable way of estimating $h$ from a vertical particle distribution $\bar\rho_p(z)$ is by fitting it to a Gaussian  form 
\begin{equation}
    \rho_p(z) = \rho_{p0} {\rm e}^{-z^2/2h^2},
\end{equation}
where $\rho_{p0}$ is an estimate for the midplane particle density, 
see \citetalias{Sengupta_Umurhan_2023} for further details.  Because the gas density is nearly constant ($=\rho_g$) we adopt the following for the midplane particle-to-gas mass density ratio, $\epsilon_0 = \rho_{p0}/\rho_g$.}


\subsection[Turbulent Stokes Numbers: \texorpdfstring{St$_L$}{St\_L} and \texorpdfstring{St$_\ell$}{St\_ell}]{Turbulent Stokes Numbers: St$_L$ and St$_\ell$}\label{sec:stokes_number}

The timescale for particles to lose all their momentum through collisions with gas molecules and become fully coupled to the gas is given by the friction (or stopping) time, $t_s$ (see Table \ref{tbl:definitions}). 
\NewEditv{
In addition to depending on the local gas density and sound speed, the particle stopping time $t_s$ varies with particle size and is often expressed in terms of a dimensionless Stokes number defined using a characteristic eddy frequency. For an eddy of size $\ell$ and wavenumber $k \equiv 2\pi/\ell$, the generalized form is}
\begin{equation}\label{eqn:stokesell}
{\rm St}_{\ell} = t_s\Omega_{\ell},
\end{equation}
with $\Omega_{\ell}$ given by
\begin{equation}\label{eqn:omegal}
\Omega_{\ell} = \sqrt{2k^3\varepsilon(k)},
\end{equation}
\NewEditv{where $\varepsilon(k)$ is defined as the total kinetic energy per unit wavenumber at wavenumber $k$.
We note that $\Omega_\ell$ is different from $\omega$ in that the latter is the spatial distribution of amplitude of vorticity while the former is the total vorticity in wavenumber space.

The largest energy-containing turbulent eddy occurs at a lengthscale $\ell = L$ (with $k_L = 2\pi/L$) that, in turn, leads to a large-eddy Stokes number}
\begin{equation}\label{eqn:stokesL}
{\rm St}_L = t_s\Omega_L,
\end{equation}
\NewEditv{where $\Omega_L$ is related to $\Omega_K$ by}
\begin{equation}\label{eqn:rprime}
R' \equiv \frac{\Omega_L}{\Omega_K} = \frac{\sqrt{2k_L^3\varepsilon(k_L)}}{\Omega_K} = 2{\rm Ro}.
\end{equation}
\NewEditv{ $R'$ is thought of as twice the large eddy's Rossby number, Ro.
In many past disk models, $\Omega_L$ has been assumed equal to $\Omega_K$, and thus ${\rm St}_L = \St_K$. However, as shown by \citetalias{Sengupta_etal_2024} and supported by the results in Table \ref{tbl:simulation}, $\Omega_L$ typically exceeds $\Omega_K$ by factors of 2–3 \REV{or even more}. This \REV{difference} 
must be accounted for when constructing effective Stokes numbers in a turbulent disk.}


\citetalias{Sengupta_etal_2024} further underscored that the true value of the turbulence intensity $\alpha$ is also subject to the same scaling by $R'$ as
\begin{equation}
    \alpha = \frac{\tilde{\alpha}}{R'} = \frac{1}{R'}\frac{U_{{\rm rms}}^2}{c_s^2},
    \label{alpha_theoretical}
\end{equation}
where $\alphat \equiv U_{{\rm rms}}^2/c_s^2$ -- often referred to as the {\textit {turbulent intensity}} -- is calculated using velocity information only with the implicit assumption of $\Omega_L = \Omega_K$.  The precise interpretation of and physical relationship between  $\alpha$ and $\alphat$
is elucidated further in \citetalias{Sengupta_etal_2024}.

Equations (\ref{eqn:rprime}) \REV{and} (\ref{alpha_theoretical}) provides a theoretical estimate of the effective $\alpha$ from the energy spectrum. This \REV{can be} 
compared against a ground-truth value, $\alpha_h$, obtained by directly measuring the particle scale height ($h$) in a simulation using a generalized form of the expression from \citet{Dubrulle_etal_1995}. i.e.,
\beq 
\frac{h^2}{H^2} = \frac{\alpha_h}{\St_K(1+\St_L^2)},
\label{alpha_measured}
\eeq
where ${\rm{Sc}}=1+\St_L^2$ is the Schmidt number \citep[e.g.,][]{Cuzzi_etal_2003,  Youdin_Lithwick_2007}. For details on how both St$_K$ and St$_L$ enter in Eqn. (\ref{alpha_measured}), see appendix E of \citetalias[][]{Sengupta_etal_2024}.

\begin{deluxetable*}{cccccccccccccc} 
\label{tbl:sim_data}
\tablecolumns{7} 
\tablecaption{List of simulations and derived quantities.}
\tablehead{ 
\colhead{Identifier} & \colhead{$h/H$} & \colhead{${\alpha_h}$}  &  \colhead{$\epsilon_{_0}$} & \colhead{$k_{_L}H$}  & 
\colhead{$R'$}  & \colhead{St$_L$} &  \colhead{$\alphat$} & \colhead{$\alpha$} & 
\colhead{$H_s/H$}& \colhead{$\delta V_0/c_s$} 
& \colhead{Ri$_\phi$} 
}
\startdata 
\vspace{-0.2cm}\\
St-01.Z-02 & 0.014 & 2.0e-6 & 1.40 & $1250 \pm 500$ & 
$3.0\pm 1.0$ &
$0.03\pm0.01$ &
$6.0^{+1.4}_{-1.3}$e-6 &
$2.0^{+0.5}_{-0.25}$e-6 &
0.018 & 0.03 & 0.310 
\\
St-02.Z-01 &  0.017 & 5.5e-6 & 0.61 & 625$\pm 75$ & $2.5^{+0.50}_{-0.75}$ & 
$0.0525\pm 0.0075$ & 
$1.5^{+0.5}_{-0.4}$e-5 &
$6.2^{+0.3}_{-0.2}$e-6 &
0.019 & 0.020
& 0.375
\\
St-04.Z-01&  0.014 &
8.4e-6 &
0.70 &
675$\pm 225$ & $3.9^{+1.1}_{-1.1}$ & 
$0.155\pm 0.045$ & 
$3.25^{+0.75}_{-0.25}$e-5 &
$8.3^{+2.7}_{-0.3}$e-6 &
0.015 & 0.022
& 0.23 
\\
\vspace{-0.2cm}\\
\hline
\vspace{-0.2cm}\\
\enddata 
\end{deluxetable*}

\subsection[Azimuthal Richardson Number \texorpdfstring{${\rm Ri}_\phi$}{Ri\_phi}]{Azimuthal Richardson Number ${\rm Ri}_{\phi}$}{\label{sec:richardson}

The Richardson number ${\rm Ri}$ is a dimensionless quantity that compares the destabilizing effect of shear with the stabilizing role of buoyancy oscillations.
In Ekman layers like these under axisymmetric settings, there are two Ri numbers that can be defined. One would be a radial Ri number, which can be used to diagnose whether the vertically sheared radial velocity field can undergo Kelvin-Helmholtz instability. Instead, we are concerned with an effective azimuthal Ri number that is essential for diagnosing the so-called Symmetric Instability 
\citep[e.g.,][]{stamper_Taylor_2017}. The azimuthal Ri$_\phi$ is here defined following the general construction promulgated by \citet{Sekiya_1998} and \citet{Chiang_2008}
\citepalias[see also][]{Sengupta_Umurhan_2023}
, in which 
\begin{subequations}
\begin{equation}\label{eqn:riphi}
    {\rm Ri}_{\phi}(z) = -\frac{\Omega_K^2z \bar\rho_p(z)}{\bar\rho_p(z) + \rho_g}\left(\frac{1}{\bar\rho_p(z)}\frac{\partial \bar\rho_p(z)}{\partial z}\right)\Bigg/\left(\frac{\partial {\cal U}}{\partial z}\right)^2.
\end{equation}
\NewEditv{In the above definition, ${\cal U}$ is the Gaussian fit to $\bar{U}_y(z)$, the radially averaged azimuthal gas velocity, computed in the same manner as $\bar{\rho}_p$ following Eq. (\ref{eq:radial_average_rhop}). The fit assumes}
\beq
{\cal U}(z) = -0.05 c_s + \delta V_{0}e^{-z^2/2H_s^2},
\eeq
\end{subequations}
\NewEditv{where $\delta V_{0}$ and $H_s$ are obtained using the same error-minimizing procedure described in Sec. \ref{sec:par-scale-height}. Within the particle layer, ${\rm Ri}_{\phi}$ is generally uniform; the values reported in Table \ref{tbl:sim_data} correspond to its midplane value.}

\subsection{Characterizing clusters and voids}}\label{Clustering_Definitions}
 
To examine the structure of the flow quantities both within and external to the voids \NewEditv{in the particle field}, we group grid points of the simulations into several different sets as defined below.
All the following sets are restricted to grid points $|z_i| < z_{\rm{lim}}$, where $z_{\rm{lim}}$ is the height below which void-filament complexes are confined (see Sec. \ref{sec:results}).  Additionally:
\par
\begin{description}
   \item[$\setV  \ \ $] The set of grid points that contain no particles.
   \item[$\setVN\ $] The set of grid points that contain no particles and whose eight surrounding neighbors also contain no particles.
   \item[$\setF\ \ $] The set of grid points that contain particles.
   \item[$\setFN\ $] The set of grid points that contain particles and whose eight surrounding neighbors are empty of particles.
\end{description}
 For example, the histogram shown in Fig. \ref{Particle_Density_St_2} is built on the set $\setF$. 
 The designations are ``f" for filament, ``v" for void, and ``n" neighbors.  Applications will follow later.
In some cases we will consider all grid points with $|z|<z_{{\rm lim}}$, which would be the union $\setF\cup \setV$.
 \par
\REV{For} given grid point $\left\{x_j,z_i \right\}$, its eight neighbors are assessed on the usual D8 indexing scheme \NewEditv{(the eight surrounding grid points)}, {\it i.e.}, 
 $\left\{x_{j},z_{i\pm 1} \right\}$, $\left\{x_{j\pm 1},z_i \right\}$, and $\left\{x_{j\pm 1},z_{i\pm 1} \right\}$.
\subsection{The Radial Distribution Function}\label{theRDF}
\REV{To quantify particle clustering, we employ the radial distribution function (RDF), $g(r)$, which measures spatial correlations in the particle distribution.} \REV{This is a widespread metric in studies of particle-gas dynamics \citep[e.g.,][]{Pan_etal_2011,Bragg_Collins_2014,Hartlep_Cuzzi_2017}.}
\REV{In three dimensions, the RDF is defined such that the mean number of particles within a spherical shell of radius $r$ and thickness $\mathrm{d}r$ around a reference particle is
\begin{equation}
\mathrm{d}N(r)=n_{p} g(r) 4\pi r^2 \mathrm{d}r,
\end{equation}
where $n_{p}$ is the mean particle number density. For a spatially random (Poisson) particle distribution, $g(r)=1$, whereas values $g(r)>1$ indicate clustering.

The RDF can in principle be computed directly from particle pair statistics by counting the number of particle pairs separated by a distance $r$. However, the computational cost of direct pair counting scales as $\mathcal{O}(N_{\mathrm{p}}^2)$ and becomes computationally prohibitive for large particle numbers $N_{\mathrm{p}}$. Instead, the RDF can also be computed from the two-point autocorrelation function of the particle density fluctuation field,
\begin{subequations}
\begin{equation}
\xi(\mathbf r)=
\left<
\delta_p(\mathbf x)
\delta_p(\mathbf x+\mathbf r)
\right>,
\end{equation}
resulting in
\begin{equation}
g(r)=1+\xi(r),
\end{equation}
with $\xi(r)$ corresponding to the autocorrelation binned by scalar distances $r$.

For the present simulations, the particle density field $\rho_p(\mathbf x)$ is first mapped onto the computational grid and normalized by the mean vertical particle density profile (equation~16) in order to remove the large-scale particle stratification. Using the particle scale height $h$ listed in Table~3, we define the normalized particle density fluctuation field as
\begin{equation}
\delta_p(\mathbf x)=
\frac{\rho_p(\mathbf x)}
{\bar{\rho}_p(z)}
-1.
\end{equation}

The autocorrelation function is evaluated efficiently in Fourier space using the convolution theorem. Specifically, we compute the Fourier transform of the density fluctuation field 
\beq
\delta_p({\mathbf k}) \equiv
\mathcal{F}\left[\delta_p(\mathbf x)\right]
\eeq
where $\mathcal{F}$ is the Fourier Transform operation,
and then go on to form the corresponding power spectral density
\begin{equation}
P_p(\mathbf k)=
\delta_p({\mathbf k}) \delta_p^{*}({\mathbf k}),
\end{equation}
where the asterisk denotes complex conjugation. The autocorrelation function is then obtained from the inverse Fourier transform
\begin{equation}
\xi(\mathbf r)=
\mathcal{F}^{-1}\left[P_p(\mathbf k)\right].
\end{equation}
\end{subequations}

Because the present simulations are axisymmetric, the correlation function is evaluated only in the radial--vertical $(x,z)$ plane. Furthermore, while the radial direction is periodic, the vertical direction is not. Direct application of Fourier methods would therefore introduce spurious correlations across the vertical domain boundaries. To avoid this artifact, we employ zero padding in the vertical direction prior to computing the Fourier-space autocorrelation.

Zero padding reduces the number of valid particle pairs contributing to the correlation at separations approaching the vertical domain size. The raw autocorrelation of the padded field is therefore weighted by the number of valid pairs at each lag, which varies with separation. Dividing by the autocorrelation of a binary mask — defined to be unity within the original computational domain and zero in the padded region — corrects for this variation and recovers an unbiased estimate of $\xi(\mathbf r)$.

The resulting two-dimensional correlation function is finally binned over scalar separations $r=|\mathbf r|=\sqrt{x^2+z^2}$ in the radial--vertical plane to obtain the RDF.
Because the simulations are axisymmetric, the two-dimensional correlation in the $(x,z)$ plane fully captures the three-dimensional particle distribution, and the resulting RDF retains its standard interpretation as a measure of scale-dependent clustering.
}
\begin{figure*}[ht]
\begin{center}
\includegraphics[width=0.7\textwidth]{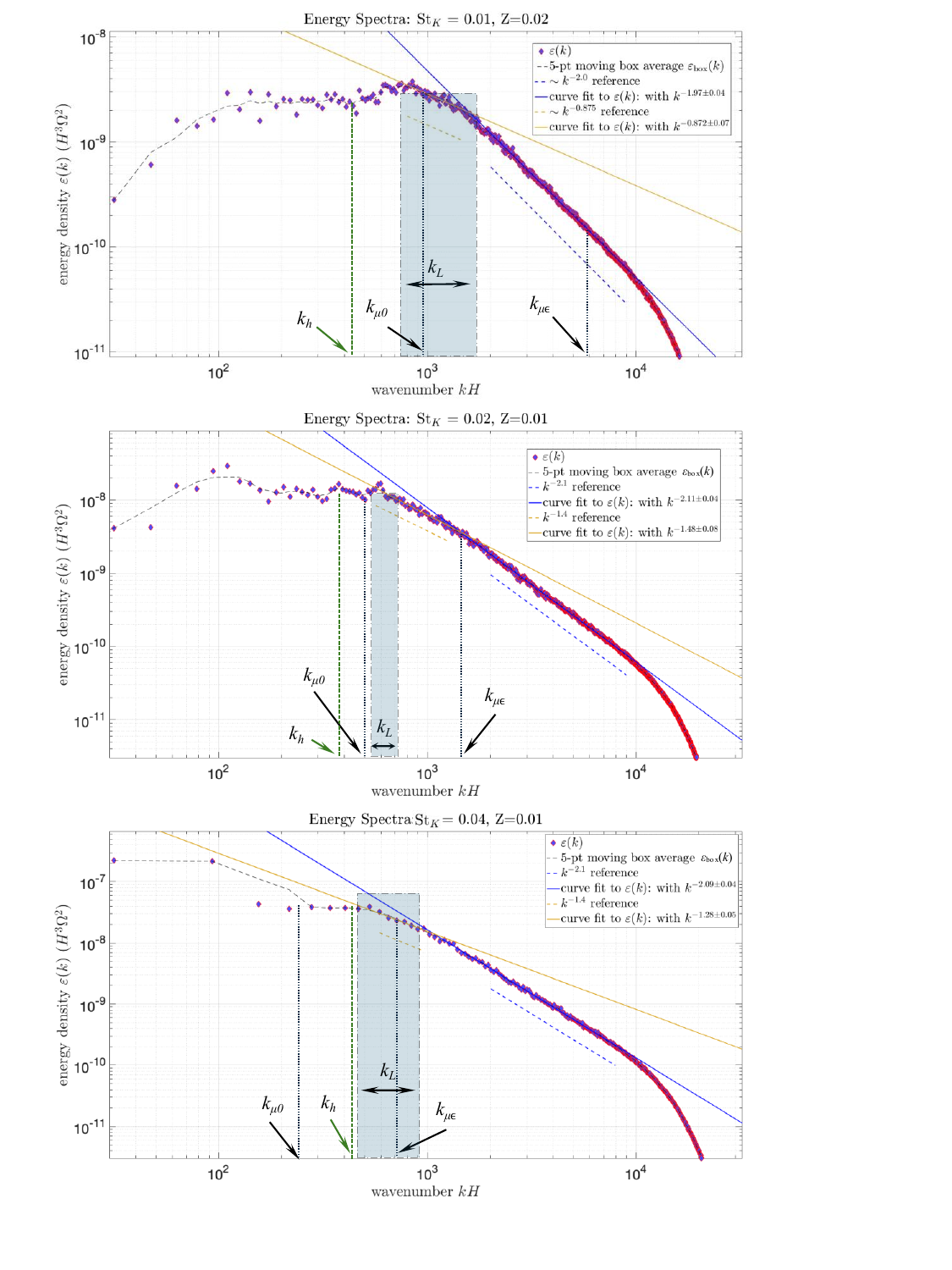}
\par
\end{center}
\vspace{-0.2in}
\caption{Specific gas kinetic energy spectra $\varepsilon(k)$ as a function of wavenumber $k$ at $\Pi = 0.05$: (top) $\St_K$ = 0.01, $Z$ = 0.02; (middle) $\St_K$ = 0.02, $Z$ = 0.01; (bottom) $\St_K$ = 0.04, $Z$ = 0.01. Reference wavenumbers shown include $k_h = 2\pi/h$ (particle scale height), the injection scale $k_L$ (band of energy input wavenumbers; see next figure), and $k_{\mu\epsilon},k_{\mu 0}$ (fastest-growing SI modes, see text). $\varepsilon_{\rm box}$ denotes a five-point boxcar average of $\varepsilon$. Two fit curves are shown: one for $k > k_L$ (revealing a clear inertial range), and one for an intermediate range generally containing the band $k_L$ the latter of which shows only qualitative scaling—possibly a broken power law. Hyperviscosity reduces power at small scales above $kH \geq 10^4$.}
\label{fig_All_Spectra}
\end{figure*}

 \begin{figure*}
\begin{center}
\includegraphics[width=0.85\textwidth]{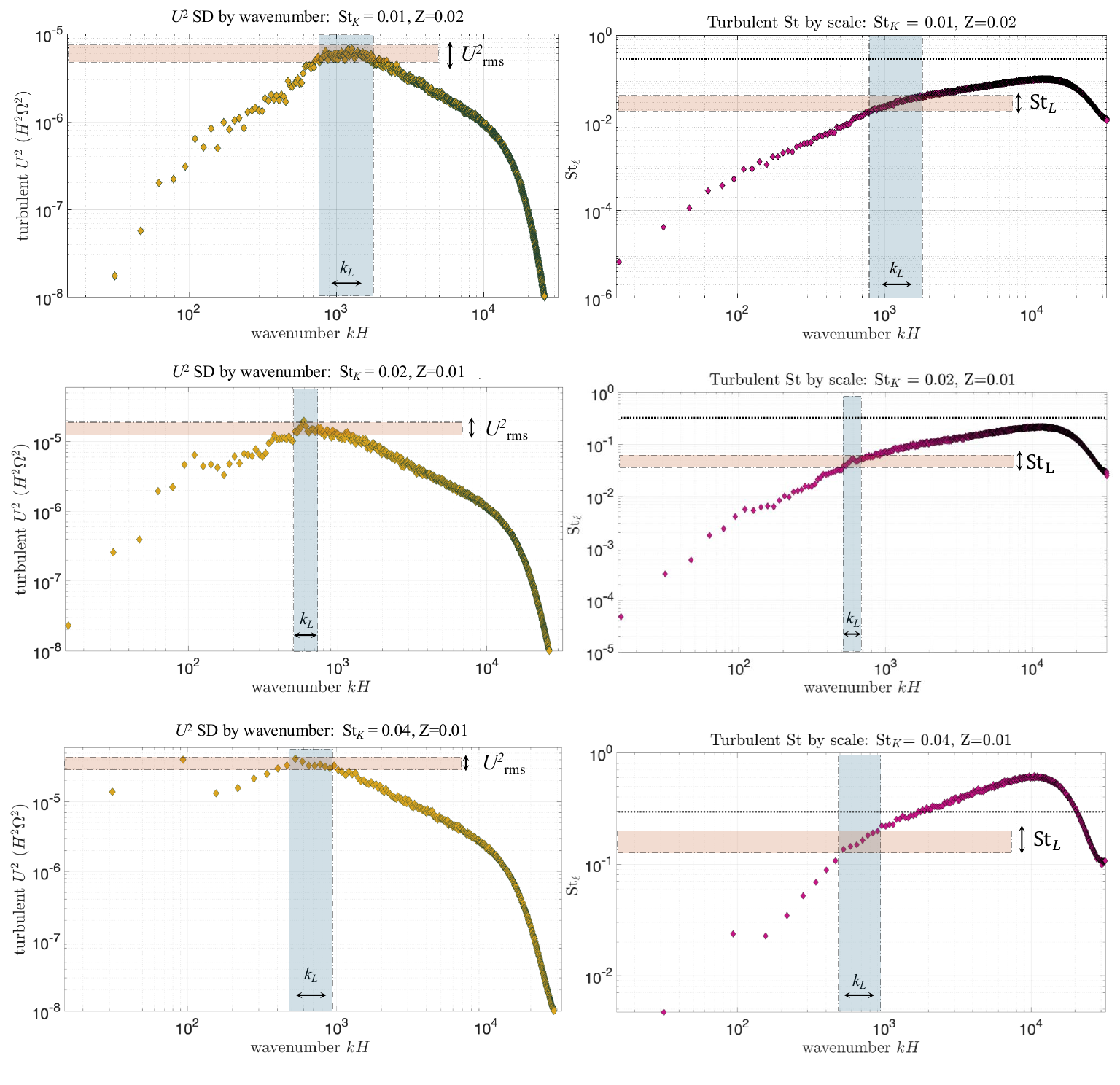}
\par
\end{center}
\vspace{-0.2in}
\caption{\underline{Left column}: SD of $U^2$ as a function of $k$. \underline{Right Column}: Effective $\St_\ell(k)$ (Eq. 19) shown for 
$\St_K$ = 0.01, $Z$ = 0.02 (top); $\St_K$ = 0.02, $Z$ = 0.01 (middle); and $\St_K$ = 0.04, $Z$ = 0.01 (bottom).  All simulations have $\Pi = 0.05$.
The range of $k_L$ is shown in the shaded light blue bands.  $U_{rms}$ is estimated as the peak of the $U^2$ SD and shown in the peach colored bands in the left column.  Similarly, the effective $\St_L$ is indicated in the corresponding horizontal bands in the right column.  The values reported in Table \ref{tbl:sim_data} are approximate averaged values shown here noting that $R' \equiv \St_L/\St_K$ (c.f., $\Omega_\ell = \St_\ell/\St_K$). $\St_\ell = 0.3$  is indicated by the black dotted line 
 in the right column panels for reference. \NewEditv{Note $\St_\ell = 0.3$ is never achieved in the upper two cases}.
}
\label{fig_U2_Rprime}
\end{figure*}

\section{Survey of results}\label{sec:results}

\subsection{Turbulence spectra}\label{turbspec}

Table \ref{tbl:sim_data} summarizes some of the quantities we derive from analyzing these simulations.
In Fig.~\ref{fig_All_Spectra}, we show the kinetic energy spectra for all simulations, all of which 
correspond to times well within the shear-turbulent phase. Each plot is annotated with the effective wavenumber associated with one particle scale height, $k_h \equiv 2\pi/h$, and the estimated radial wavenumber of the fastest growing mode, $k_{\mu\epsilon}$, given by a generalization of the the resonance criterion of \citet{Squire_Hopkins_2018a} corresponding to the radial wavenumber of the fastest growing SI mode (see also \citealt{Umurhan_etal_2020}):
\beq \label{eqn:kmu}
k_{\mu\epsilon} \approx \frac{(1+\epsilon_{_0})^2}{2\St_K \Pi},
\eeq
which holds reasonably well in the $\St_K \leq 0.1$ regime, \NewEditv{and is well resolved}.  
We input values of $\epsilon_0$ as quoted in Table \ref{tbl:simulation}, and further define $k_{\mu 0} \equiv k_{\mu\epsilon}(\epsilon = 0)$. We also indicate the range of wavenumbers corresponding to the estimated injection scale, $k_L$. 

Figure~\ref{fig_All_Spectra} shows that an inertial range with scaling $\varepsilon(k) \sim k^n$, where $-2 \gtrapprox n \gtrapprox -2.1$, emerges in all simulations for $kH > 2000$ and persists until numerical dissipation from hyper-viscosity sets in (typically near $kH \approx 10^4$). We also include a suggestive fit line spanning the approximate injection range, $k_L$, up to $kH \approx 1500$. In this range, the slope depends on the simulation parameters but is generally flatter than that of the shorter-wavelength inertial range.  \NewEditv{We note that the inertial-range slope in these dynamically restricted axisymmetric simulations {\it is not expected} to match the Kolmogorov value of $-5/3$ characteristic of three-dimensional, homogeneous, isotropic turbulence \citepalias[see][]{Sengupta_Umurhan_2023}. Recognition of this deviation from the classical value is central to our subsequent conclusions regarding the prevalence of TC in these simulations.}
\par

The left column in Figure~\ref{fig_U2_Rprime} shows the \REV{closely related} \NewEditv{spectral decompositions (SD)} of $U^2 = 2k\varepsilon_k$ for the modeled combinations of $\St_K$ and $Z$. The range of $k_L$ indicated in each spectrum of Fig.~\ref{fig_All_Spectra} is qualitatively estimated by identifying the wavenumber band where $U^2$ is maximized, which we associate with $U^2_{\rm rms}$.  
The right column of Fig.~\ref{fig_U2_Rprime} shows the corresponding $\St_\ell$. For each case, we indicate the putative range of $\St_L$ associated with the $k_L$ band discussed above. The corresponding $R'$ values, listed in Table~\ref{tbl:simulation}, typically fall within the range of 3 to 4 (or, 2 to 5 including uncertainty bar ranges) consistent with the findings for driven turbulence discussed in \citetalias{Sengupta_etal_2024}.

In the St-01.Z-02 run, 
\NewEditv{the favored SI development scale $k_{\mu\epsilon}$ lies well within the inertial range and is well separated from the putative injection scale $k_{\rm L}$}.
We also note that $k_L$ is confined within a particle scale height, approximately between a factor of 2 and 3 ({\it i.e.}, $k_L \approx 2$–$3 \times k_h$). \NewEditv{That is, the large eddy scale is a few times smaller than the particle layer thickness}. The same qualitative trends hold for the St-02.Z-01 run, except the range in $k_L$ is much narrower and lies closer in scale to $k_h$. Similarly, the fastest growing mode $k_{\mu\epsilon}$ is shifted toward the start of the inertial range but remains well within it. By contrast, in the St-04.Z-01 run, the situation is different: the length scale of the fastest growing mode appears to lie inside the $k_L$ range, with $k_h$ situated near the lower bound of that range. This suggests that this particular parameter combination marks a transition in the character of the turbulent dynamics, \REV{which we discuss further in Section 5}.

The value of $\alpha$ inferred from the energy spectrum (Eq. \ref{alpha_theoretical}) compares reasonably well with the measured value $\alpha_h$ derived from the particle scale height. As summarized in Table \ref{tbl:sim_data}, $\alpha_h$ generally falls within the error bars of $\alpha$. For the St-02.Z-01 run, $\alpha_h = 5.5 \times 10^{-6}$ is approximately 10\% lower than the corresponding estimate derived from spectral analysis.  We observe that had \REV{the commonly adopted} $\tilde\alpha$ been \REV{calculated from the observed $h$,} 
the resulting 
values would be a factor of $R'$ higher than that of the directly measured value $\alpha_h$ -- we also revisit this in Section \ref{sec:discussion}.

\begin{figure*}
\begin{center}
\includegraphics[width=0.5\textwidth]{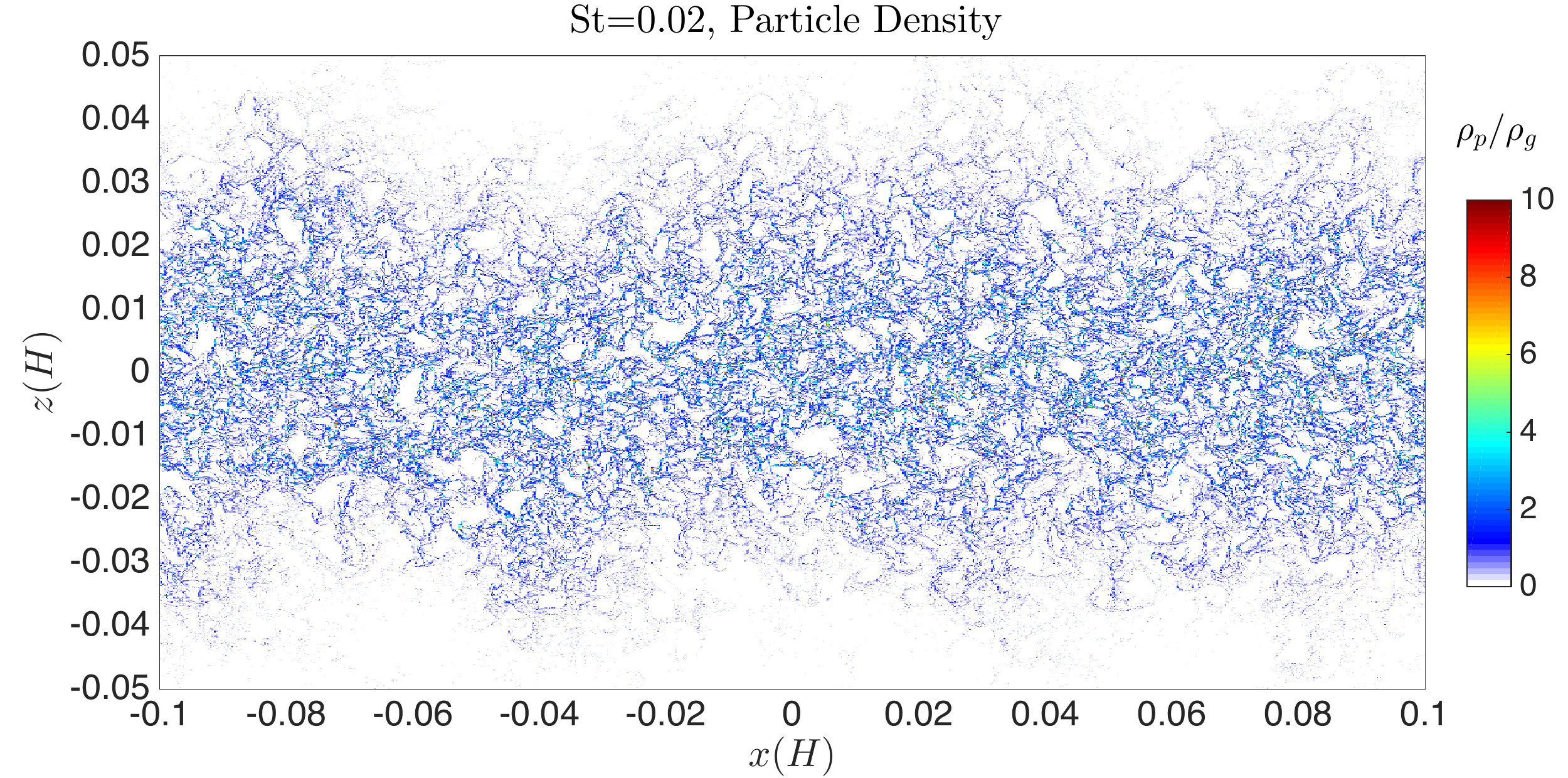}
\includegraphics[width=0.202\textwidth]
{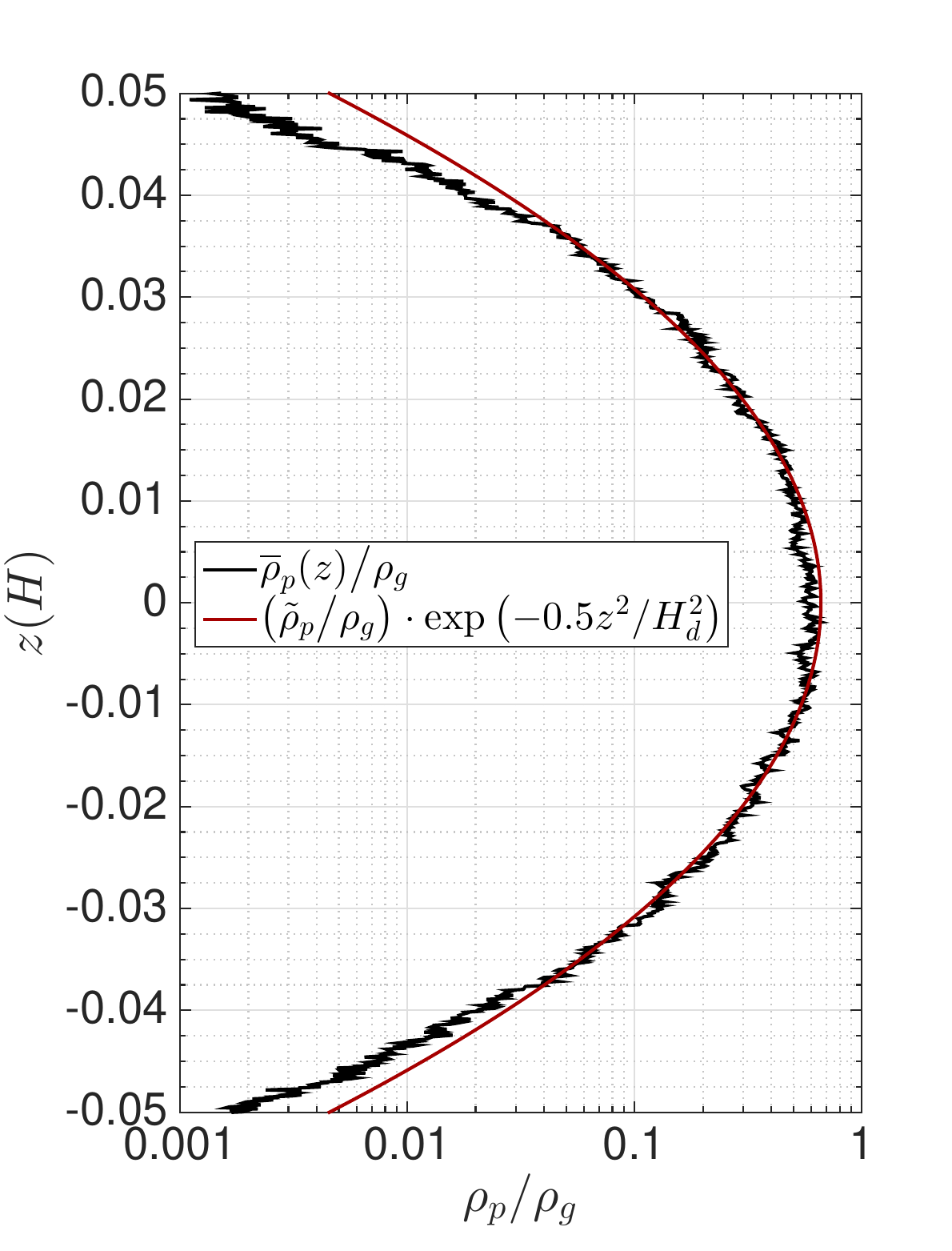}
\ \ \ 
\includegraphics[width=0.27\textwidth]{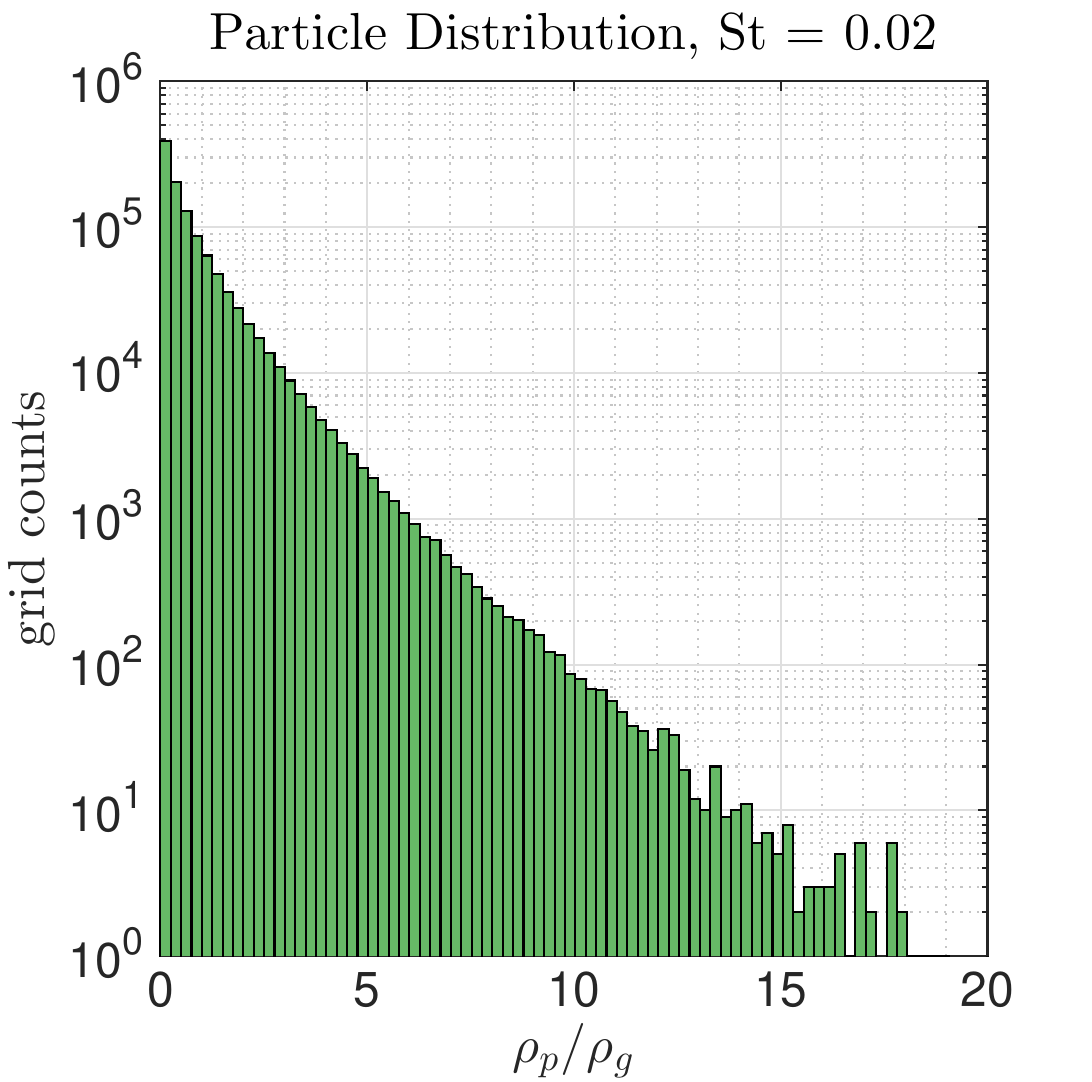}
\\
\par
\end{center}
\vspace{-0.2in}
\caption{Particle density for $\St_K = 0.02$.  The left panel shows a typical map of particles with its characteristic void-filament complexes threading generally everywhere $|z|<z_{{\rm lim}} \approx 0.032 H$.  The middle panel shows $\overline\rho_p(z)$, together with a Gaussian fit with scale height $h \approx 0.016 H$.  The right panel is a histogram of the particle density restricted to the grid set ${\cal X}^{(f)}$ and exhibiting an exponential character.}
\label{Particle_Density_St_2}
\end{figure*}

\begin{figure}
\begin{center}
\leavevmode
\includegraphics[width=0.45\textwidth]{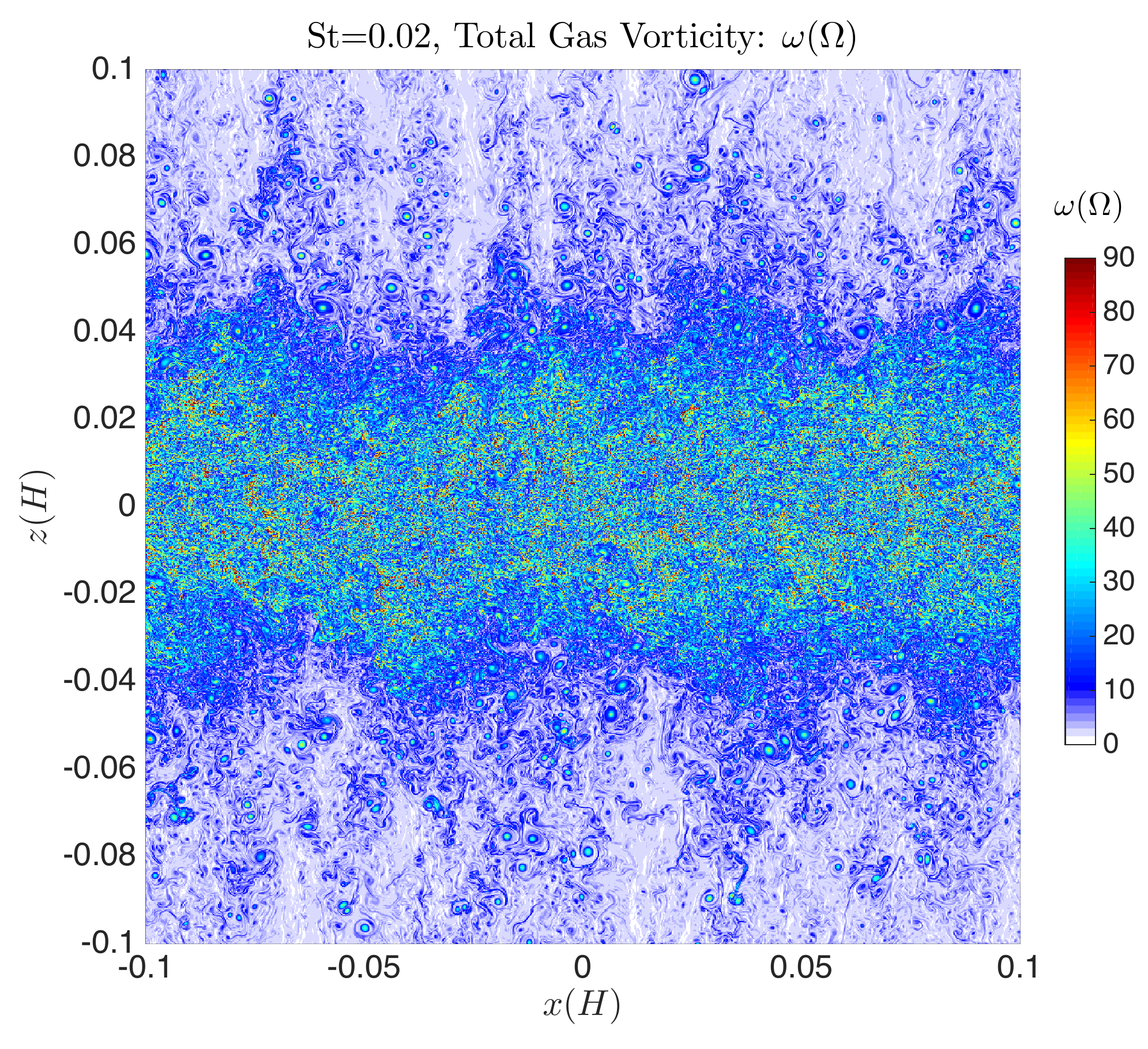}
\par
\end{center}
\caption{Total vorticity $\omega$ plot for simulation St-02.Z-01.  The characteristic radially wavy structure typical of fully developed SI filaments is weakly manifest. 
  Note the change in intensity at around the $z=\pm z_{{\rm lim}}$.
 The region shown is not the full box domain, where $L_x = L_z = 0.4 H$.  The color coding for vorticity is in units of local disk rotation frequency $\Omega_K$.
  }  
\label{St_2_gas_vorticity}
\end{figure}

A central tenet of TC is that particle distributions exhibit peak intermittency (sometimes referred to as ``lacunarity”) at scales where $\St_\ell = \order{0.3}$ \citep[e.g., see discussion around Figs. 10 and 13,][]{Hartlep_etal_2017}. The right column of Fig. \ref{fig_U2_Rprime} shows the SD of $\St_\ell$ as a function of $k$. Among the three simulations, only St-04.Z-01 resolves the length scale where $\St_\ell = 0.3$ within the inertial range, corresponding to a length scale near $\ell/H \approx 0.003$, or about 30 grid points.\footnote{A sharp downturn in $\St_\ell$ appears at scales of 10–12 grid points, consistent with the onset of hyperviscosity. Turbulent dynamics are not expected to be accurately captured on smaller scales \citepalias{Sengupta_Umurhan_2023}.} In contrast, the $\St_\ell(k)$ profiles for the other two runs never intersect the critical $\order{0.3}$ threshold within the resolved range, indicating that the 
\NewEditv{maximum intermittency (clumping) is never achieved}.
Nevertheless, as we show below, signatures of 
TC 
still emerge, even in these under-resolved simulations.

\subsection{Spatial distributions of vorticity and particle concentration}\label{spatdist}

The left panel of Fig. \ref{Particle_Density_St_2} shows the particle distribution for simulation St-02.Z-01 that typifies the fine scale particle clumping that manifest in all of our experimental runs once the particles have settled into a midplane quasi-steady turbulent state. The emergence of large fluid patches free of particles is robust  \NewEditv{-- and whose presence has been reported in previous published models like that of \citet[][see their Figure 1]{Yang_etal_2017}}.  These particle-free zones will be referred to as {\it voids} henceforth. 
\REV{The voids are bounded by} \REV{streams of particles we call} 
\REV{{\it filaments}. 
The void-filament complexes generally appear to be confined to $|z| \le  0.03 H$ which approximately corresponds roughly to 2 particle scale heights (see Table \ref{tbl:sim_data}). 
To keep our discussion flexible, in what follows we consider the statistics of enclosed voids for all values of $|z| \le z_{{\rm lim}}$ with the upper bound being generally less than $2h$.
For example, the middle panel of Fig. \ref{Particle_Density_St_2}  shows $\overline\rho_p(z)$, where we fit a value of $h \approx 0.016 H$, demonstrating that the transition from filament bounded voids into open voids occurs roughly at 2$h$.}  
The right panel shows a histogram distribution of the number of grids containing particles of a given total particle density {\it greater than zero}.  
The histogram indicates an exponential distribution.

 \begin{figure*}
\begin{center}
\includegraphics[width=0.63\textwidth]{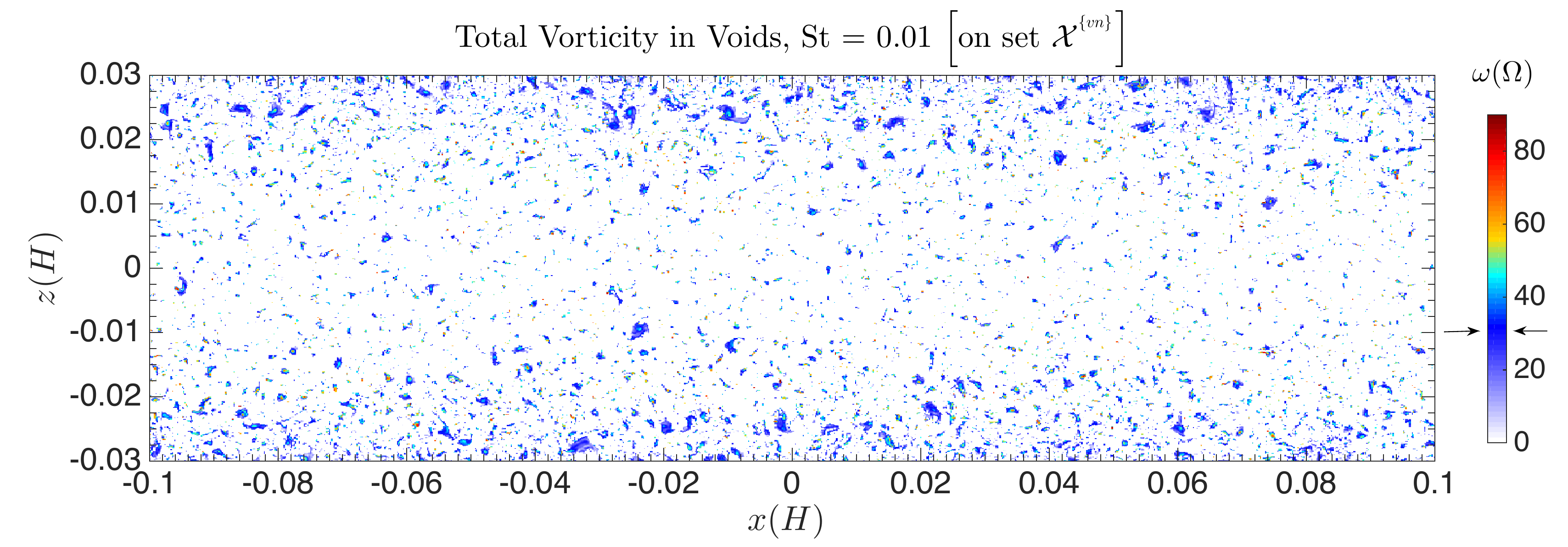}
\ \ \ \
\includegraphics[width=0.27\textwidth]{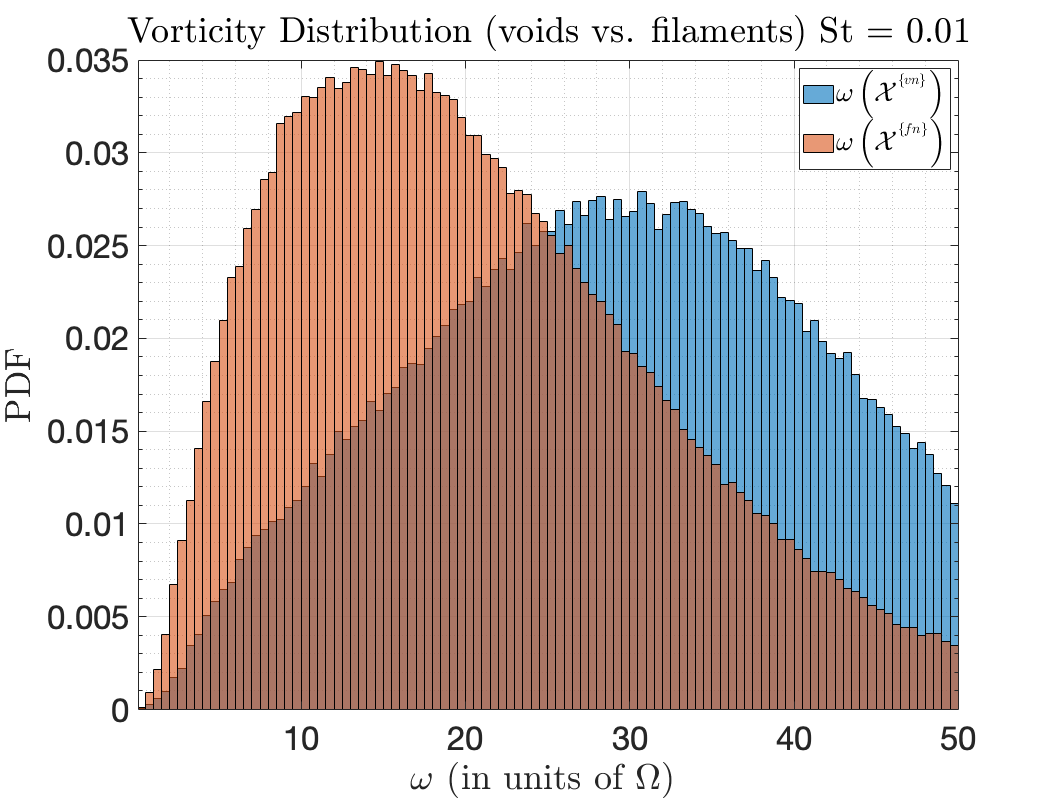} \ \ \ \\
\includegraphics[width=0.63\textwidth]{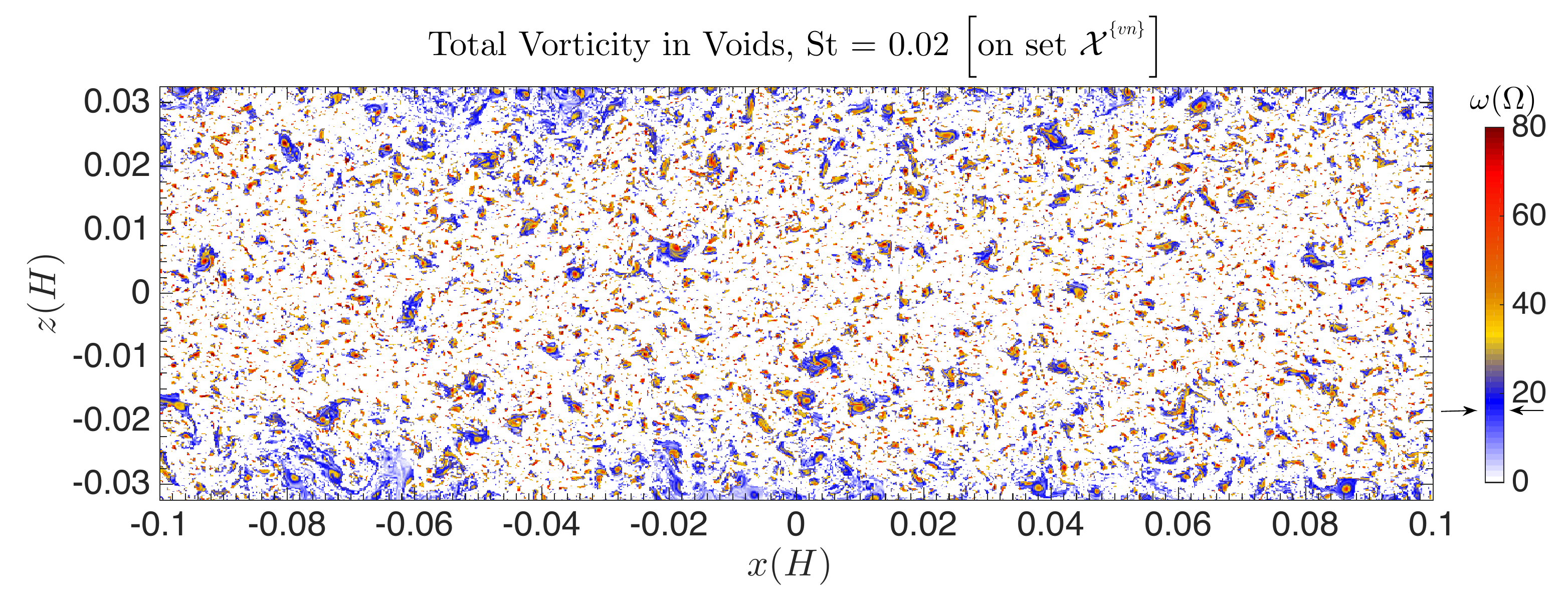}
\ \ 
\includegraphics[width=0.28\textwidth]{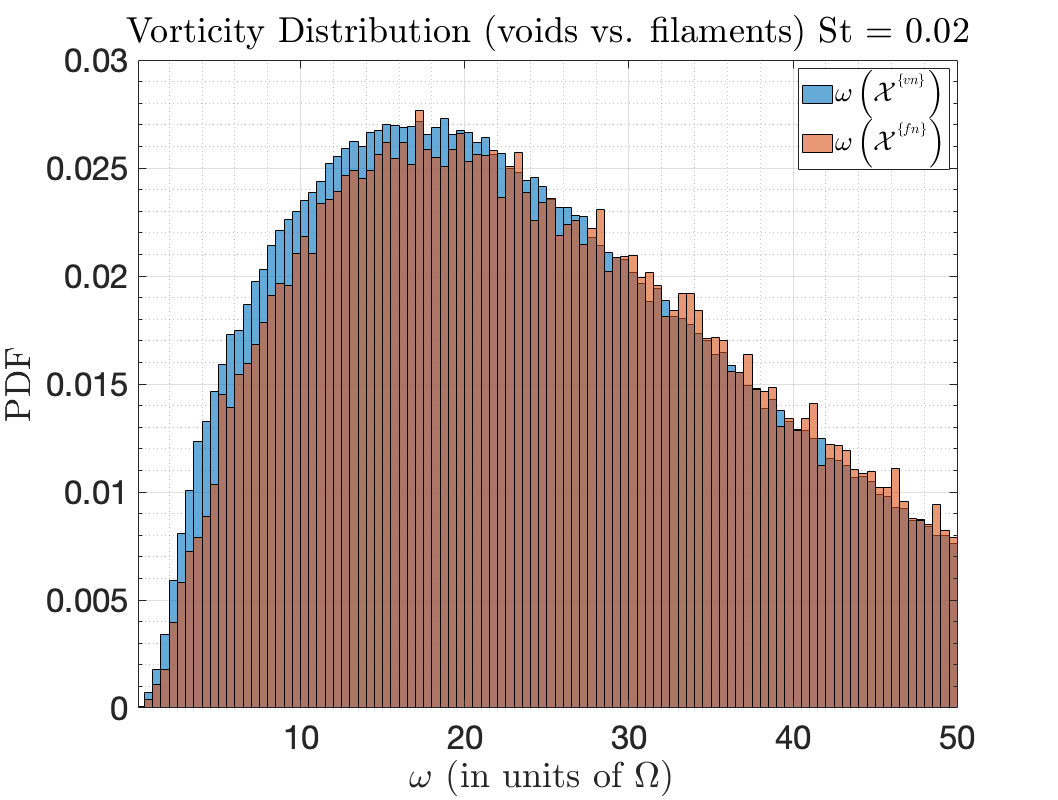}\\
\includegraphics[width=0.63\textwidth]{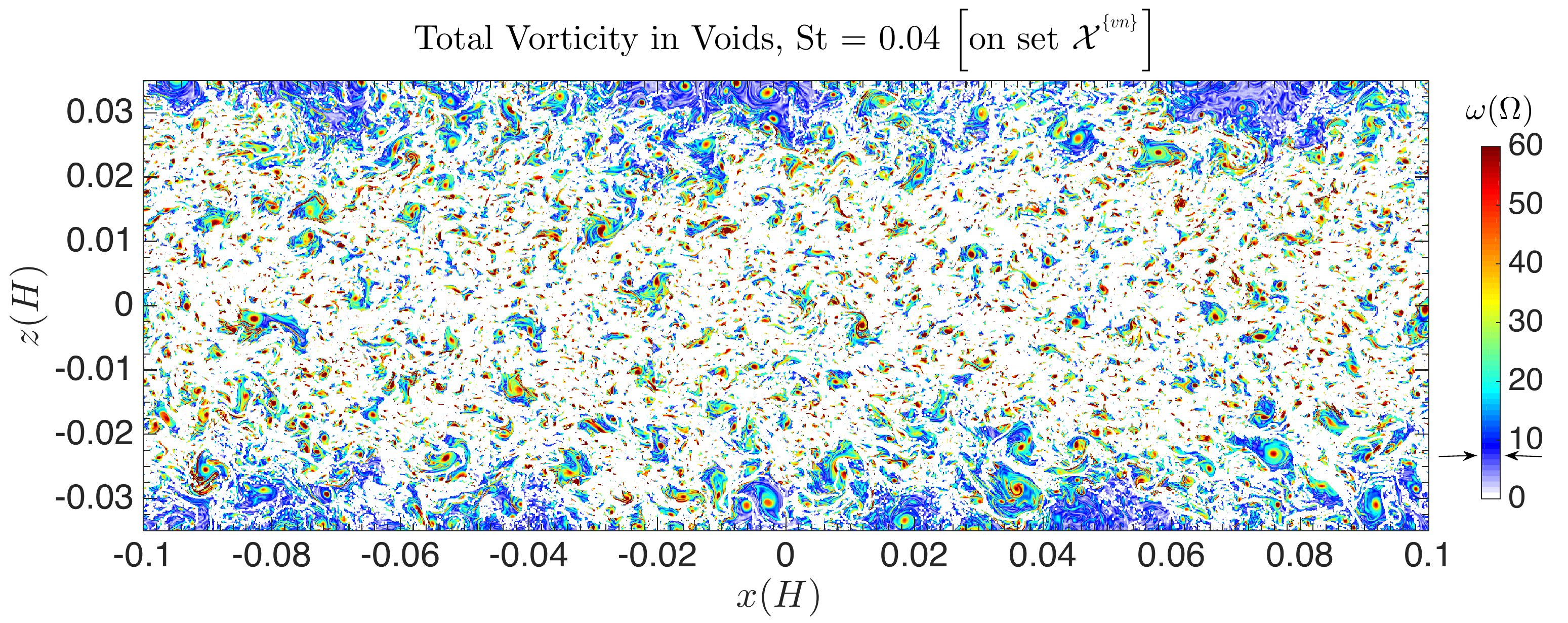}
\ \ 
\includegraphics[width=0.27\textwidth]{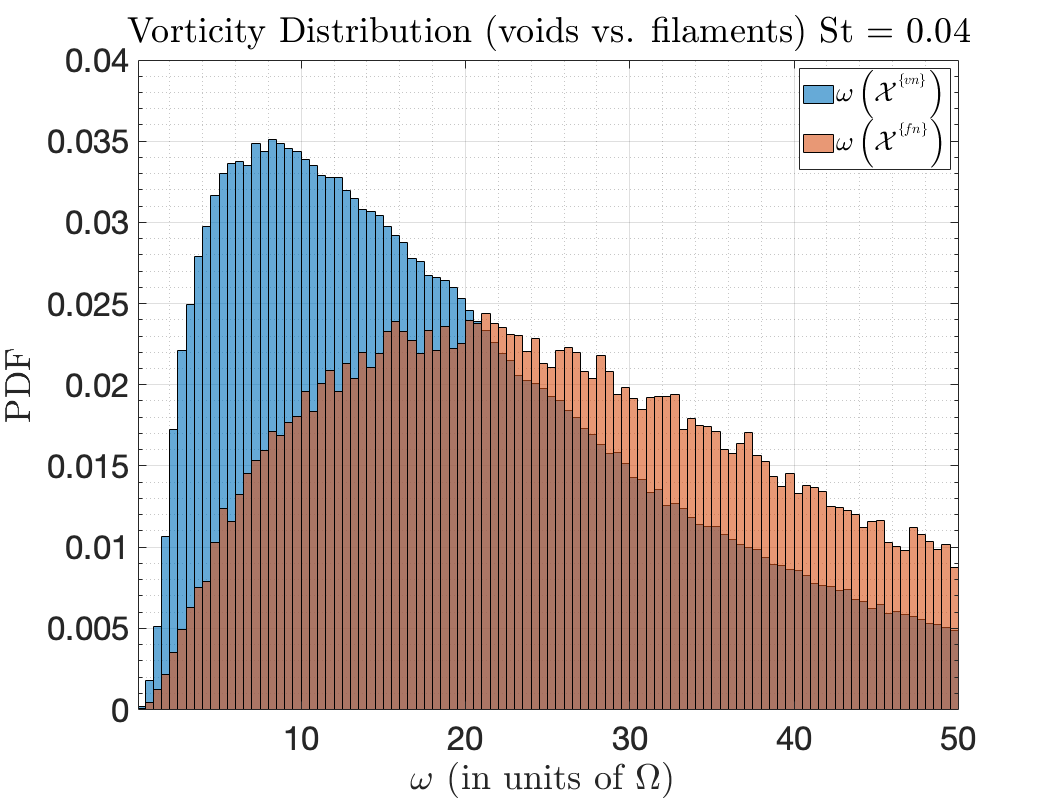}\\
\par
\end{center}
\vspace{-0.2in}
\caption{\underline{Left column}:
Vorticity maps of void regions ($\setVN$) within the particle layer for $\St_K = 0.01$ (top), $\St_K = 0.02$ (middle), and $\St_K = 0.04$ (bottom). The vorticity in filament regions is masked (white). \underline{Right column}:
Vorticity PDFs within the particle layer, shown for voids ($\setVN$, blue) and filaments ($\setFN$, orange), for each corresponding $\St_K$ value in the left column. \REV{These distributions assume $z_{{\rm lim}} = 0.03H$.} Arrows on each \REV{vorticity color} bar mark the peak vorticity within voids ($\setVN$). All distributions are approximately log-normal. Note: The peak vorticity ($\omega_{\rm peak}$) in filaments varies little with $\St_K$, while the void region peak shifts significantly. }
\label{Composite_New}
\end{figure*}

 \par
In Figure \ref{St_2_gas_vorticity} we display the total vorticity for the 
simulation 
shown in Figs. \ref{fig_U2_Rprime} and \ref{Particle_Density_St_2}.  The range in $\omega$ goes from almost zero to over $90 \Omega_K$.  A stark change in the character and magnitude of $\omega$ can be seen around the putative \REV{$|z|\approx \pm 2h$ level, which corresponds to the transition from filament-enclosed voids (i.e., $|z| \le 2h$) to open voids ($|z| \ge 2h$).  Within the particle layer, $\omega$ appears generally disordered}, interspersed with regions of structured vorticity (which we examine further) while away from the layer we witness mainly structured, putatively dipolar vortices weakly concentrated along vertically oriented pillars.   Such \REV{layered} bursts of gas flow from the particle layer are typical in
axisymmetric calculations of this sort
\citepalias[e.g., as extensively discussed Sec. 3 of][\citealt{Li_Youdin_2021} also report this phenomenon]{Sengupta_Umurhan_2023}.
Qualitatively speaking it can be seen that $\omega$ is larger within the layer.  These qualitative trends follow for the St-01.Z-02 and St-04.Z-01 simulations.
\par
In 
Figure \ref{Composite_New}, we show the gas vorticity for each simulation, focusing on midplane regions where particles are primarily located. The intensity of the vorticity is rendered only at grid points belonging to the void set $\setVN$—those that contain no particles and whose D8 neighbors are also particle-free (see Sec. \ref{Clustering_Definitions}). All other regions are masked in white. These void regions exhibit coherent, structured vortices. To the right of each vorticity map, we display the distribution of scalar vorticity $\omega(x,z)$, restricted either to $\setVN$ or to the filament set $\setFN$, which consists of all grid points identified as part of particle filaments.

An important \REV{qualitative} feature can be identified from the histograms presented for $\omega$ in the right column of Fig. \ref{Composite_New} for both the voids and the filaments. The vorticity in filaments peaks between $15 - 20 \Omega_K$ across the $\St_K$ used in the simulations. However, the vorticity in the voids shifts significantly towards lower values \REV{(and larger scales)} as $\St_K$ is increased. The peak vorticity $\omega_{{\rm peak}}$ for $\St_K=0.01$ is around $30 \Omega_K$, while $\omega_{\rm peak}$ is less than $\sim 10 \Omega_K$ for $\St_K=0.04$. The $\omega_{{\rm peak}}$ in $\setVN$ in each case is marked by a double arrow on the colorbar in the corresponding surface plots on the left column. 

 \begin{figure*}
\begin{center}
\includegraphics[width=0.32\textwidth]{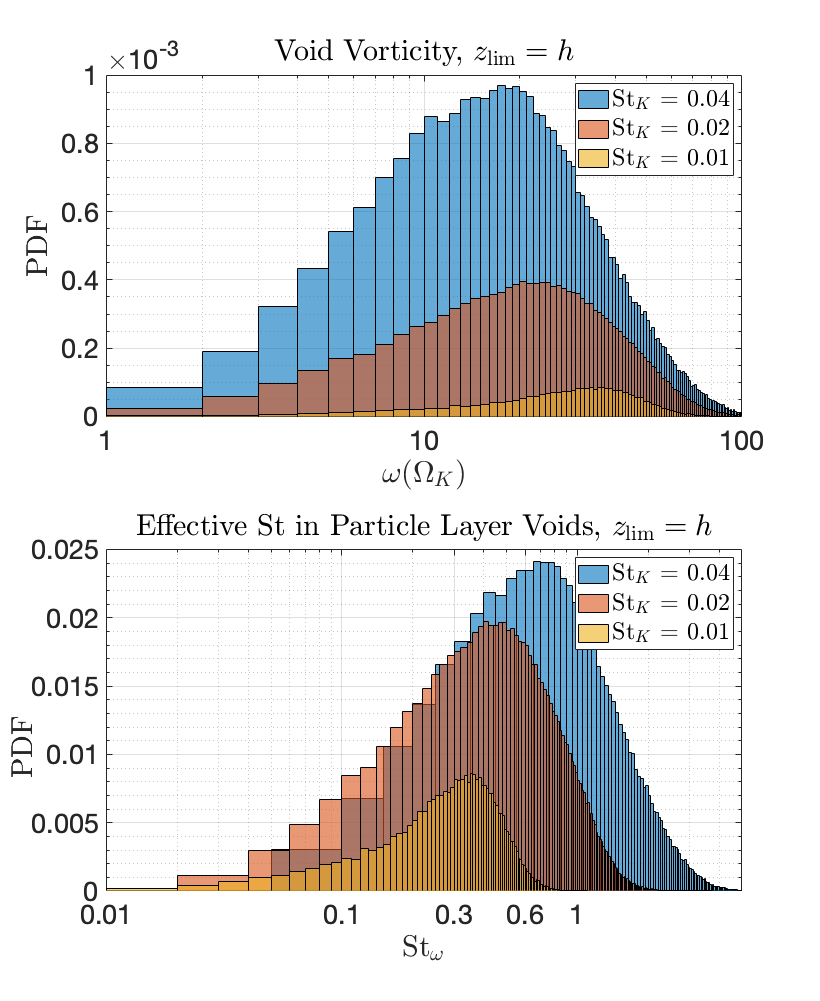}
\includegraphics[width=0.32\textwidth]{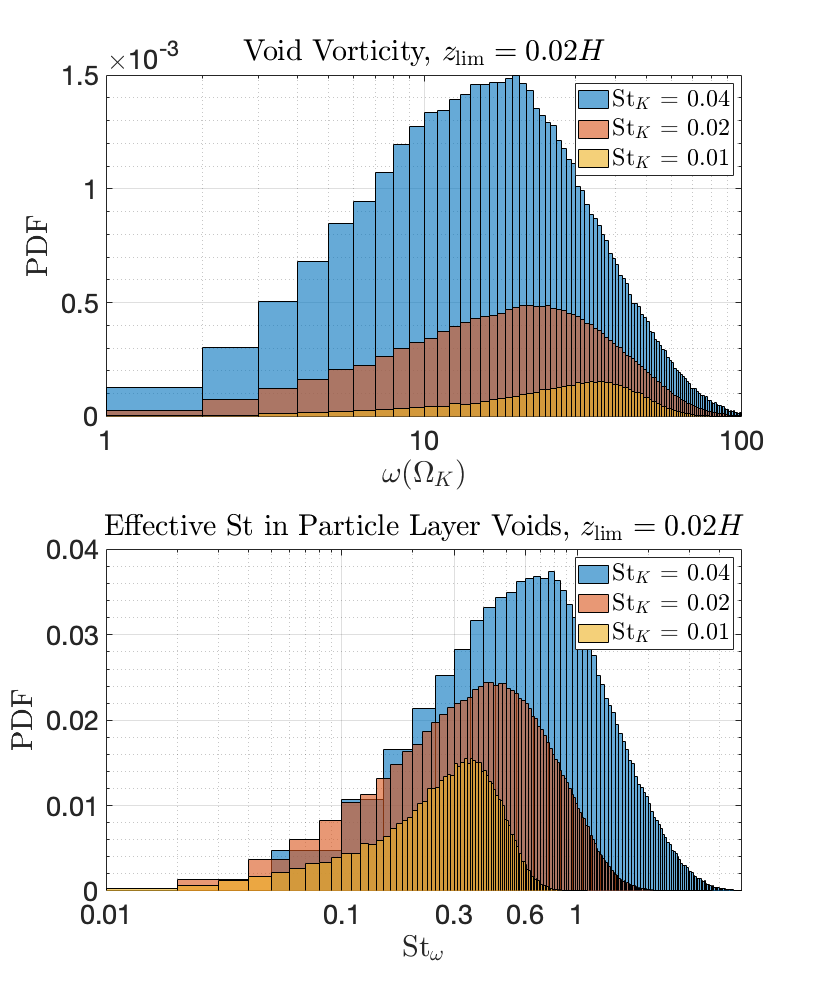}
\includegraphics[width=0.32\textwidth]{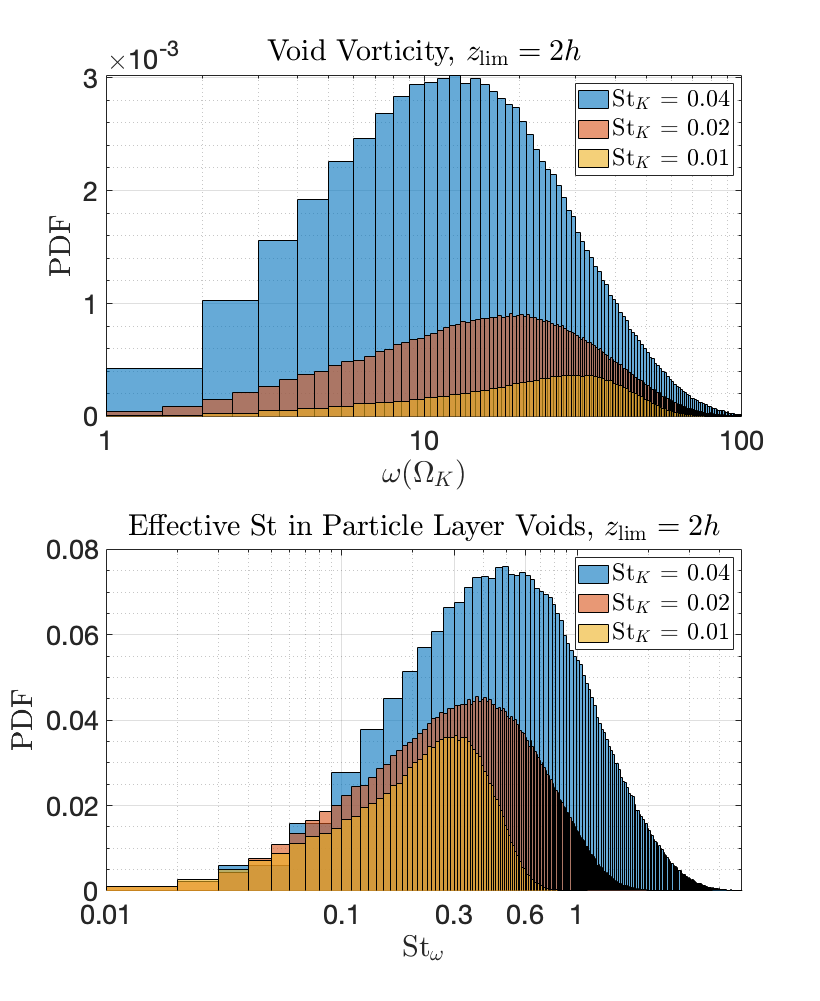}

\par
\end{center}
\vspace{-0.2in}
\caption{\REV{Vorticity \REV{PDF} in voids.
(Top row): \REV{PDF} of the void vorticity drawn from the void grid-point set (${\cal X}^{(vn)}$) for each simulation (see also previous figure).
(Bottom row): \REV{PDF} of the effective Stokes number, $\St_\omega$, \REV{in voids} for the three simulations. From left to right we show increasing $z_{{\rm lim}}$: (left) $z_{{\rm lim}} = h \approx 0.016 H$, (middle) $z_{{\rm lim}} = 0.02H$, (right) $z_{{\rm lim}} = 2h$.
The left and middle panels \REV{of the bottom row suggest} convergence: for each $\St_K$ simulation the peak satisfies $0.4 < \St_\omega({\rm peak}) < 0.75$. The right panel shows \REV{all peaks} drifting toward smaller $\St_\omega$, suggesting contamination by vorticity from outside the particle layer.}}

\label{Three_St_Collapse}
\end{figure*}

\begin{figure*}
\begin{center}
\includegraphics[width=0.99\textwidth]{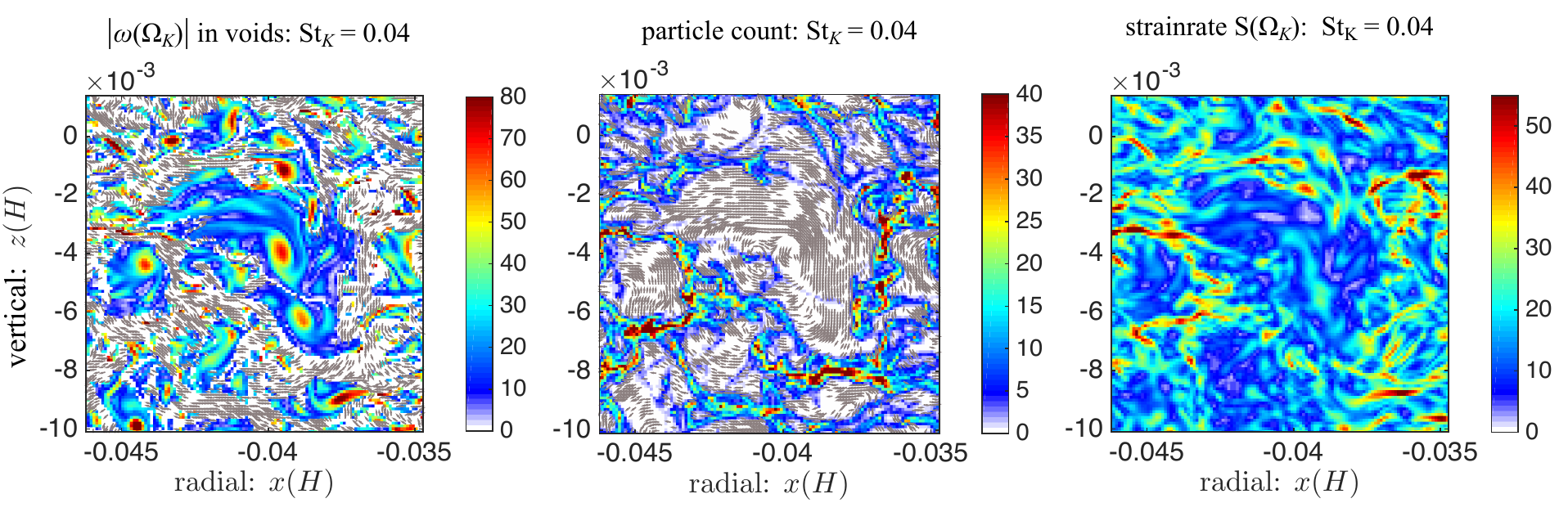}
\par
\end{center}
\vspace{-0.2in}
\caption{\NewEditv{
A subsection of the full domain for simulation St-04.Z-01 is shown. 
Color maps depict the absolute vorticity in particle voids, 
$\omega$, normalized by $\Omega_K$. 
White regions mark particle locations. 
\REV{Enlargement of} the left panel \REV{reveals} particle velocity vectors (grey) over the void vorticity field. 
The middle panel presents particle counts, with gas velocity vectors drawn 
in the intervening voids. 
The right panel displays the corresponding gas strain rate, $S/\Omega_K$.}
Note the correlation between the particle count and the strain rate, and the anti-correlation between the particle count and the local vorticity. This is direct evidence of the fact that TC is operative in the system.
}
\label{fig:cross_comparison}
\end{figure*}

\REV{In Figure \ref{Three_St_Collapse}, the underlying vorticity distributions are shown in the top row and closely resemble those of the voids in the right column of \ref{Composite_New}, where $z_{{\rm lim}} = 0.03H \approx 2h$ for all simulations considered. We also show the probability distribution function (PDF) of the effective Stokes number, $\St_{\omega}$, evaluated over the void set $\setVN$ for all three simulations and for three choices of $z_{{\rm lim}}$ (bottom row). Recall that $\St_{\omega} = t_s \omega$ is defined using the local vorticity and is therefore distinct from St${_\ell}$ introduced in Sec. \ref{sec:definitions}.}
\REV{Weighting this vorticity by the corresponding $\St_K$ to form St${_\omega}$ on $\setVN$ yields a distribution whose peak, $\St_\omega$(peak), consistently lies in the range $0.4$–$0.7$ (see Table \ref{tbl:peak_St_omeg}). This behavior remains robust for $z_{{\rm lim}} \le 0.02H$. For $z_{{\rm lim}} = 2h$, however, $\St_\omega$(peak) shifts to smaller values, $\sim 0.3$–$0.35$. We attribute this shift to contamination of the vortex statistics by vorticity originating outside the particle layer, which cannot be cleanly excluded in the present analysis.}

\begin{deluxetable}{c|cccc} 
\label{tbl:peak_St_omeg}
\tablecaption{$\St_\omega$(peak) estimates for several $z_{{\rm lim}}$ }
\tablehead{ 
\colhead{$\St_K$}  & \colhead{$z_{{\rm lim}} = 0.01H$}  
& \colhead{$z_{{\rm lim}} = h$} & 
\colhead{$z_{{\rm lim}} = 0.02H$} & 
\colhead{$z_{{\rm lim}} = 2h$}
}
\startdata 
0.01 & 0.35 & 0.35 & 0.35 & 0.31 
\\
0.02 & 0.45 & 0.44 & 0.43 & 0.38 
\\
0.04 & 0.70 & 0.72 & 0.75 & 0.48 
\\
\hline
\enddata 
\end{deluxetable}

It is also evident from Fig. \ref{Three_St_Collapse} that the \REV{vorticity} distribution becomes wider as St$_K$ increases. Furthermore, for St$_K=0.01$, the distribution is skewed towards smaller values of St$_{\omega}$, whereas it becomes more symmetric for St$_K=0.04$. We discuss this feature in more detail later in Section \ref{TCalways}. 

In Figure \ref{fig:cross_comparison}, a small section within the particle layer is extracted from the entire simulation domain of the St-04.Z-01 run to demonstrate \REV{more clearly} the ongoing operation of TC. The left panel shows the vorticity of the voids $\omega(\Omega_K)$ 
in units of the local Keplerian frequency $\Omega_K$. The particle count is shown in the same domain in the middle panel. The right panel shows the \REV{local} strain-rate $S(\Omega_K)$. A cross comparison between $\omega(\Omega_K)$ 
and particle count shows that the void regions with coherent vortices are selectively avoided by the particles, and the particle count reaches its maximum in pockets of the domain just outside the regions containing the vortex structures. Similarly a \REV{direct} correlation between particle count and strain-rate  can be observed when comparing the middle and the right panel. The high strain-rate regions match nicely with those having high particle counts.

\subsection{Radial distribution functions}\label{RDFs}

\begin{figure*}[ht]
\begin{center}
\includegraphics[width=0.95\textwidth]{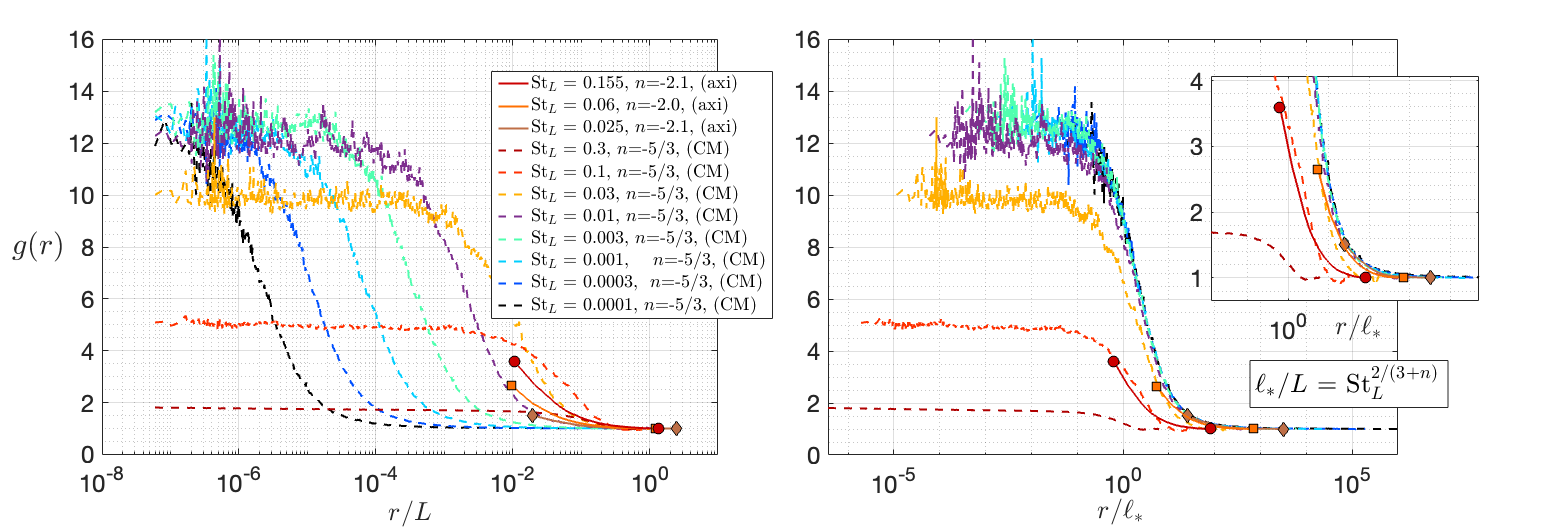}
\par
\end{center}
\vspace{-0.2in}
\caption{
\REV{
        \REV{Radial distribution functions, $g(r)$, for the present axisymmetric simulations compared with the 3D Kolmogorov cascade-model ("CM") predictions of \citetalias{Hartlep_etal_2017}. The RDFs are expressed in terms of the effective large-eddy Stokes number, $\St_L$, corresponding to each simulation's $\St_K$ (see text).}
        \underline{Left}: $g(r)$ as function of $r/L$.
        Solid lines denote the simulations, while dashed lines show the cascade-model predictions, which exhibit a characteristic S-shaped dependence with the rise occurring at scales that depend strongly on Stokes number
        \underline{Right}: same $g(r)$ but plotted as function of $r/\ell_*$, where $\ell_*$ is the scale at which the scale-local Stokes number $St_\ell$ becomes unity (Eq. \ref{def:ellstar}).
        Model curves for small $St_L$ collapse when scaled in this way, while larger Stokes numbers follow the same shape but experience less clustering at large scales and therefore reach smaller asymptotic values as $r\to0$.
        Curves from the axisymmetric simulations reproduce the cascade-model trends well but sample only a limited range of the predicted curves.
        For the two smallest-Stokes-number cases, the simulations do not probe scales at or below $\ell_*$, and the inflection point of the S-curve therefore remains unresolved.
}
}
\label{fig:RDF_v1}
\end{figure*}

\REV{ 
We conclude by presenting the RDFs (Sec. \ref{theRDF}) from our simulations. In the left panel of Figure~\ref{fig:RDF_v1}, \(g(r)\) is shown for the three axisymmetric runs with separation \(r\) expressed in integral-scale units \(L=2\pi/k_L\) (see Table~\ref{tbl:simulation} for \(k_L\)). Curves are labeled by \(\St_L = R'\St_K\), defined in Eqs.~(\ref{eqn:stokesL})–(\ref{eqn:rprime}) and listed in Table~\ref{tbl:simulation}.
For quick reference the correspondances are:
$\St_K = 0.01\leftrightarrow \St_L = 0.03$, 
$\St_K = 0.02\leftrightarrow \St_L = 0.05$, and
$\St_K = 0.04\leftrightarrow \St_L = 0.15$. 
For comparison, we include the \citet[][H17 hereafter]{Hartlep_etal_2017} cascade–model results for $g(r)$—an inexpensive, validated statistical proxy for full 3-D particle–turbulence simulations \citepalias[see][and references therein]{Hartlep_etal_2017}.

}



\REV{
\textit{Interpreting $g(r)$}: $g(r) \rightarrow 1$ when $r$ is sufficiently large recovering a normalized unclustered Poisson-like distribution. If a single clustering scale is present, $g(r)$ will exhibit an ``S-curve", remaining nearly flat for large $r$, then rising toward an inflection point as $r$ approaches the nominal clustering scale $r_c$, then proceeding to flatten out as $r$ continues toward smaller $r$.\footnote{In general, if multiple clustering scales exist, successive S-curve onsets appear as r decreases, each marking an additional characteristic scale.}
TC predicts that this nominal value of $r_c \approx \ell_*$, where $\ell_*$ corresponds to the lengthscale where $\St_\ell = \order 1$.  }

\REV{\citetalias{Hartlep_etal_2017} analyzed 3D high-resolution particle-gas turbulence data from \citet{Bec_etal_2010} to develop a refined cascade model for the clustering of inertial particles.
In this framework, clustering arises through scale-dependent concentration amplification across successive cascade levels.
From their analysis, it was shown that the strength of intermittency in particle density fields depends on the scale-local Stokes number St$_{\ell}$.
Following} 
the theory for isotropic homogeneous turbulence of \citet{Kolmogorov_1941}, \citetalias{Hartlep_etal_2017} defined St$_{\ell}$ as 
\begin{subequations}
\begin{equation}
    {\rm St}_{\ell} = \frac{t_s}{t_{\ell}} = {\rm St}_L\left(\frac{\ell}{L}\right)^{-2/3},
    \label{def:H17_St_ell}
\end{equation}
where $L$ is the energy injection scale. From their new refined cascade model of particle clustering, 
\REV{\citetalias{Hartlep_etal_2017} derive a prediction for the radial distribution function (their Fig. 13), which exhibits the characteristic S-shaped dependence of turbulent clustering. Their findings show that the RDF rises most rapidly at scales for which St$_{\ell}=\order 1$, before approaching an asymptotic clustering level at smaller scales.}

\REV{It is critical to realize that the} use of Eq. (\ref{def:H17_St_ell}) in our simulations—specifically the exponent on $\ell$—is not 
\REV{appropriate} because our axisymmetric runs do not follow a Kolmogorov scaling in their kinetic energy spectra like we see in Fig. \ref{fig_All_Spectra} \citepalias[and also see][]{Sengupta_Umurhan_2023}. To allow for generalization, we instead write
\beq
\St_\ell = \St_L \left(\frac{\ell}{L}\right)^{(-3-n)/2}, \qquad {\rm for} \qquad  \ell \le L,
\eeq
\end{subequations}
where $n = -5/3$ for isotropic Kolmogorov turbulence, and $n \approx -2$  for the axisymmetric simulations considered here
(see Fig. \ref{fig_All_Spectra}).


\REV{
The TC prediction that strong clumping onsets at scales
$\ell$ satisfying
$\St_\ell = \order 1$ naturally motivates casting the radial dependence of the RDFs in terms of the length scale
\beq
\ell_* \equiv \St_L^{2/(3+n)} L. \label{def:ellstar}
\eeq
The right panel of Fig.~\ref{fig:RDF_v1}
rescales separations by $\ell_*$, collapsing the left-panel RDFs onto a pattern governed primarily by $\St_L$.
\REV{As first noted by \citetalias{Hartlep_etal_2017}, plotting $g(r)$ against $r/\ell_*$ yields nearly identical curves with the same small-$r$ asymptotic value as $r/\ell_* \to 0$ for $\St_L < 0.01$. This functional dependence is also remarkably similar to the theoretical prediction of \citet{Zaichik_Alipchenkov_2009}, although their model predicts a smaller asymptotic value. For $\St_L > 0.01$, the small-$r$ asymptote progressively weakens. Our axisymmetric RDFs agree with the cascade-model prediction in the onset location, with the rise in $g(r)$ beginning at approximately the same $r/\ell_*$ for corresponding $\St_L$ values (right inset of Fig.~\ref{fig:RDF_v1}).}
}

\REV{However, none of our axisymmetric runs reaches the expected small-$r/\ell_*$ asymptote in $g(r)$, and the low-St cases ($\St_K=0.01,0.02$) do not attain the S-curve inflection evident in the cascade models. 
}
\REV{The $\St_K=0.04$ case shows the strongest development of high-amplitude intermittency but still does not fully traverse the predicted S-curve, indicating that the full cascade signature is not yet resolved.}


\REV{Focusing more on the $\St_K=0.04$ simulation, we consider the scale-local Stokes number computed directly from the measured turbulent spectrum, St$_{\ell} = {\rm St}_K \left(\Omega_{\ell}/\Omega_K\right)$ (bottom right panel of Fig. 2), rather than using the scaling relation for St$_\ell$ used above. The value St$_{\ell} \sim 0.3$, corresponding to the scale of strongest concentration amplification in the \citetalias{Hartlep_etal_2017} cascade model, lies at $kH \sim 2000$, well within the inertial range. Moreover, the inertial range extends to $kH \sim 10^4$, providing substantial scale separation below this point. The simulation therefore resolves both the scales of strongest clustering amplification and a sizable fraction of the scale range over which clustering continues to build, even if the full asymptotic regime is not achieved. Consistent with this picture, the measured clustering exhibits the characteristic S-curve behavior discussed above, providing strong evidence for turbulent clustering in this run.}

\section{Discussion}\label{sec:discussion}

\subsection{TC is always present}\label{TCalways}

\REV{The presence of TC is demonstrated in our simulations in several ways.}
\REV{First, in} Fig. \ref{fig:cross_comparison} \REV{(Sec. \ref{spatdist})} the particles systematically populate regions of high strain-rate. A cross-comparison between the left and the middle panel of Fig. \ref{fig:cross_comparison} shows that particles avoid regions of voids containing the coherent vortical structures. A distinct void region is observed in the color-map for the particle count coinciding with a region where vorticity reaches its peak in the left panel. Similarly, a correlation is observed  between the particle concentrations (middle panel) and the strain-rate map. A central tenet of TC is that particles are systematically ejected from the regions of high vorticity into adjacent regions with high strain-rate -- a feature that clearly manifests itself in the simulation with $\St_K=0.04$.

\REV{Second,} the operation of TC in the simulation is further supported \REV{in Sec. \ref{spatdist} by agreement of} the void length scales \REV{with those associated with} St$_{\ell} = 0.3$. From the bottom right panel of Fig. \ref{fig_U2_Rprime}, St$_{\ell} = 0.3$ roughly occurs where $kH \sim 2000$, which corresponds to a length-scale $\sim 0.003 H$. \REV{Meanwhile,} an eyeball estimate for the length-scales of the voids that contain the coherent vortex structure in the left panel of Fig. \ref{fig:cross_comparison} lie between $0.0015 H$ and $0.003 H$. This remarkable consistency between the two values of $\ell$ assessed independently supports the robust operation of TC \NewEditv{throughout the settled midplane layer}. 

\REV{Third, the RDF analysis (Sec. \ref{RDFs}, see Fig. \ref{fig:RDF_v1}) agrees very well with predictions of TC, and indeed suggest our simulations have not yet} 
captured the {\it maximum} intermittency in the particle density field. For St$_K=0.04$ the length-scale corresponding to St$_{\ell}=0.3$ lies somewhere in the middle of the inertial range. However, we can see from the right column of Fig. \ref{fig_U2_Rprime} that for St$_K=0.02$ and $0.01$, the St$_{\ell}=0.3$ line (the horizontal dotted line) does not intersect the spectra, signifying that the resolutions we used in simulations St-01.Z-02 and St-02.Z-01 are not enough to capture the maximum particle clustering \REV{for these $\St_K$ values.}  


\REV{On a final note,} we 
turn our attention to the vorticities of the voids containing coherent vortices and the related St$_\omega$ distribution as depicted in Figure \ref{Three_St_Collapse}.  \REV{For the three simulations spanning different $\St_K$, the distribution of $\St_\omega$—defined by \REV{scaling} $\St_K$ by the local vorticity $\omega$ of the void set $\setVN$—collapses onto a narrow range of peak values. These span from $\sim 0.4$ for $\St_K = 0.01$ to $\approx 0.7$ for $\St_K = 0.04$. This range lies close to the preferred turbulent Stokes number, $\St_\ell \simeq 0.3$, associated with particle clustering, with the understanding that $\St_\ell$ and $\St_\omega$ are not identical by definition.} \REV{Indeed, we believe that this preference for $\St_\ell \simeq 0.3$ in the most intermittent regions is another manifestation of the RDF results shown in Figure \ref{fig:RDF_v1}.} 
\par
\REV{
An important issue, \REV{however, was} raised during the review process \REV{which we address here}. 
We have assumed that the void vortices play a direct role in producing the observed particle concentration. However, clustering may depend not only on $\St_\omega$, which compares the particle stopping time to the local eddy turnover time, but also on the time required for particles to traverse an eddy. If particles cross a coherent structure much faster than the eddy can deflect them, their interaction with that structure may be limited, even when $\St_\omega$ is putatively favorable for such deflection to occur. This concern is especially relevant for the $\St_K=0.04$ simulation and motivates examining particle velocities measured directly from the turbulent flow. \textit{We have tested this possibility and find that it does not apply to the simulations examined here.} The methods employed supporting this conclusion are detailed in Appendix \ref{sec:appendixUpg}.
}


\subsection{The driving mechanism.}\label{sec:driving_mechanism}

Generation of turbulence by and in a particle-laden midplane layer of the protoplanetary disk remains a complex question.  Under conditions where no other turbulence generating linear instabilities are operative (e.g., like the VSI, COV, MRI),  the three mechanisms, KHI, SymI, and SI, have been proposed to be simultaneously active there \citepalias[][]{Sengupta_Umurhan_2023}. Of the three, SI is generally thought to be the primary mechanism for particle clustering and planetesimal formation.  Is this really the case? 


\REV{We use}
our simulations in detail to assess the viability of SI as a turbulence-driving mechanism at relevant scales. Figure \ref{fig_All_Spectra} presented the kinetic energy spectra for all three simulations, with the estimated wavenumber of the fastest growing SI mode ($k_{\mu\epsilon}$; Eq. \ref{eqn:kmu}) marked in each case. Also indicated are the driving scales, $k_L$, inferred from the spectral peaks in $U_{\rm rms}^2$ (Fig. \ref{fig_U2_Rprime}, left column).

In the $\St_K=0.04$ run, the fastest-growing SI mode ($k_{\mu\epsilon} H \sim 730$) overlaps the energy injection scale. But for the $\St_K=0.01$ and $0.02$ cases, $k_{\mu\epsilon} H$ increases significantly—reaching $\sim 5800$ and $\sim 1300$, respectively—well into the inertial range and far from the driving scale peaks, which respectively occur at $kH \sim 1100$ and $\sim 600$. This strongly indicates that the fastest growing SI modes do not drive turbulence in these two simulations. The likely alternative drivers are linked to the Ekman flow structure in the layer \citep{Dobrovolskis_etal_1999}, processes like SymI and KHI \citepalias{Sengupta_Umurhan_2023}.
\REV{Processes like the VSSI are also possibilities \citep[][]{Ishitsu_etal_2009,Lin_2021},
but the exact dynamical nature of these processes, and how they compare to and differ from Ekman layer instabilites, remains to be elucidated.}


To evaluate whether the SI is even \textit{operative} at its fastest-growing mode, we compare its predicted growth rate to the eddy turnover frequency at the corresponding scale. At any scale $\ell$, the dominant process is expected to be the one with the shorter characteristic timescale (i.e., higher frequency).

Thus, the eddy turnover frequency $\Omega_{\ell}$ at any length-scale $\ell$ and corresponding wavenumber $k=2\pi / \ell$ can be assessed from Eqn. (\ref{eqn:omegal}). Following this, the eddy turnover frequency at $k=k_{\mu\epsilon}$ are approximately $7.10\Omega_K , 4.10 \Omega_K$ and $4.15 \Omega_K$ for St$_K=0.01, 0.02$ and $0.04$ respectively, calculated using values extracted from Figure \ref{fig_U2_Rprime}.\footnote{Specifically, we read off the 
value of $\St_\ell(k_{\mu\epsilon})$ from the corresponding graphs found in the right column of Fig. \ref{fig_U2_Rprime} and divide that value by $\St_K$ to get $\Omega_\ell(k_{\mu\epsilon})/\Omega_K$.}

These turbulent disruption rates should be compared with the growth rate of the fastest-growing SI mode, which depends on whether $\epsilon$ exceeds or falls below unity. In the $\St_K = 0.01$ simulation with $Z=0.02$, the midplane value is $\epsilon \approx 1.4$. The fastest-growing mode lies in the so-called asymptotic regime,\footnote{For the SI, the fastest-growing mode occurs in the $k_z \gtrapprox k_{\mu\epsilon}$ limit at $k_x \approx k_{\mu\epsilon}$. We refer to this as the {\emph{asymptotic regime}}
\citep[see also][]{Umurhan_etal_2020}.}  
in which the growth rate is
\begin{subequations}
\begin{equation}
\sigma_{_{\mathrm{SI}}} \approx \sqrt{\epsilon - 1} \, \Omega_K,
\end{equation}
for $\St_K \ll 1$ with $\epsilon > 1$
\citep[][]{Youdin_Goodman_2005, Squire_Hopkins_2020}. For the $\St_K = 0.01$ simulation, we estimate $\sigma_{_{\mathrm{SI}}} \approx 0.65 \, \Omega_K$, which is ten times weaker than the eddy overturn frequency at that scale ($\sim 7 \, \Omega_K$). Growth rates are slower for all other $k_x$ and $k_z$ values that deviate from this asymptotic regime.  We also thus treat these $\sigma_{_{\mathrm{SI}}}$  estimates as the overall upper bound for the SI in any given simulation with given parameter values $\epsilon, \Pi, \St_K$.

From Table \ref{tbl:simulation}, we find that the midplane values of $\epsilon$ for the other two simulations with $Z=0.01$ fall below unity. In this regime, the SI growth rate becomes explicitly Stokes-number-dependent. The fastest-growing mode, again in the asymptotic regime, has a growth rate given by
\begin{equation}
\sigma_{_{\mathrm{SI}}} \approx
\frac{2 + 3\epsilon (1 - \epsilon)}{(\epsilon + 1)(4 - 3\epsilon)} \, \St_K \, \Omega_K.
\end{equation}
\end{subequations}
This empirical expression agrees well with the asymptotic growth rates reported in the $\St_K \ll 1$ and $0\ll \epsilon < 1$ limit by \citet[][see their Figure 3 and Figure 2, respectively]{Youdin_Goodman_2005, Youdin_Johansen_2007}.\footnote{We also note that \citet{Squire_Hopkins_2018a} demonstrate that in what we call the asymptotic limit with $\epsilon \ll 1$, $\sigma_{{\rm SI}} \propto 
\sqrt{\epsilon}\St_K\Omega_K$.}

Applying this to the $\St_K = 0.02$ and $0.04$ simulations yields SI growth rates of $\sigma_{_{\mathrm{SI}}} \approx 0.016 \, \Omega_K$ and $0.033 \, \Omega_K$, respectively—both much slower than 
eddy overturn frequencies at the corresponding length scales. The injection-scale overturn rates ($\Omega_L$) are also significantly faster than these SI growth rates. Together, these comparisons apparently indicate that the SI is unlikely to be the dominant driver of turbulence in the $\St_K = 0.02$ and $0.04$ runs with $Z=0.01$. For $\St_K = 0.01$ and $Z=0.02$, however, the SI might be acting in concert with another instability, such as the SymI, whose growth rate is bounded by \citepalias{Sengupta_Umurhan_2023}
\begin{equation}
\sigma_{_{{\rm SymI}}} \le
\sqrt{2\left(\frac{1}{\mathrm{Ri}_\phi} - 1\right)} \, \Omega_K.
\end{equation}
Using the derived values of $\mathrm{Ri}_\phi$ from Table \ref{tbl:simulation}, we estimate
$\sigma_{_{{\rm SymI}}}/\Omega_K \approx 2.1,\ 1.8,\ 2.6$ 
for $\St_K = 0.01,\ 0.02,\ 0.04$, respectively. These estimates are broadly comparable to the large eddy turnover frequencies for each run ($\Omega_L/\Omega_K = R'$; see once again Table \ref{tbl:simulation}). While this does not prove that the SymI is the primary turbulence driver across these three simulations analyzed, it stands out as the most plausible candidate -- particularly for the $\St_K = 0.02$ and $0.04$ simulations.

\REV{
For the $\St_K=0.01$ simulation, a concerted action of SI and SymI—akin to the VSSI \citep{Lin_2021}—may be operative. However, the VSSI's fastest–growing modes lie deep within the inertial range we resolve. At the estimated injection scales, both viscous and inviscid VSSI growth rates fall in the range $0.2\,\Omega_K \text{ to } \Omega_K,$ at least a factor of three weaker than the eddy–overturn frequencies measured in our runs \citep[cf. growth rates in Figs. 3 and 15 of][in contrast to the values of $\Omega_L$ reported in Table \ref{tbl:simulation}]{Lin_2021}. Moreover, VSSI power is concentrated near the top of the particle layer, whereas our simulations exhibit turbulent power distributed throughout the full layer, e.g., as might a visual inspection of Fig. \ref{St_2_gas_vorticity} would certainly attest.\!\footnote{\REV{In \citetalias{Sengupta_Umurhan_2023}, we hypothesized that VSSI may represent a not-yet fully characterized amalgam of SymI and SI, tending toward SymI as $\St_K\to 0$, with the SI disappearing entirely in that limit. This interpretation remains conjectural.}}

}

\subsection{Numerical considerations regarding clustering physics at the small scales}\label{sec:clustering_physics}

In Sec. \ref{sec:driving_mechanism}, we have discussed why \REV{we believe} SI is not the primary driver of turbulence in the particle-laden midplane layer, at least for the Stokes number range used in this paper, \NewEditv{that is most realistic for nebula applications}. It is also important to identify the mechanism that clusters particles at the small scale, something more directly related to the question of planetesimal formation. SI has been the preferred mechanism for planetesimal formation for the last decade or more. 
The discussion in Sec. \ref{sec:driving_mechanism} elucidates that at relevant scales, SI is weak and overwhelmingly suppressed by the local eddies at the layer-turbulence injection scales. Hence, it is quite unlikely that the small scale particle dynamics and the clustering at those scales are at all influenced by SI, but instead most likely a direct consequence of the robust operation of TC, \REV{as elaborated upon in Section \ref{TCalways}.} 

\REV{As discussed in Sec. \ref{RDFs}, attaining maximum intermittency in the particle density field—and hence maximizing the likelihood of TC-driven clustering to densities exceeding the Roche threshold—requires, at minimum, resolving the $\St_{\ell}=\order{1}$ scale. Moreover, as the RDF analysis presented in Sec. \ref{RDFs} suggests, resolving even smaller length scales may \REV{actually} be necessary.} 
Thus, \REV{detecting} the formation of planetesimal forming bound clumps at the midplane of protoplanetary disks by TC in numerical models is resolution dependent. 

Interestingly, there is evidence of this effect already present in the literature. For example, \citet{Li_Youdin_2021} found no strong clumping for St$_K=0.01$ with a dust-to-gas mass ratio $Z < 0.02$ (2\% of the solar metallicity) with a resolution of $1280$ grids per gas scale height. However, \REV{when} doubling resolution with the same domain size as \citet{Li_Youdin_2021}, \citet{Lim_etal_2025a} revealed strong clumping for the same value of St$_K$ with $Z=0.013$ and $0.01$. Although no turbulent energy spectra were provided by \citet{Lim_etal_2025a} we conjecture that by increasing the resolution, \citet{Lim_etal_2025a} resolved scales closer to St$_{\ell}=0.3$, or even exceeding it, revealing elevated small scale clustering reaching beyond the local Roche density.  

This identification of the small-scale clustering mechanism in the simulations reported here is particularly important in the context of planet formation, since $\St_K \lesssim {\rm few}\times 0.01$ is the expected Stokes-number range for particles available for planetesimal formation, as indicated by growth models \citep{Estrada_etal_2016, Estrada_etal_2022}, meteoritic evidence \citep[][and references therein]{Simon_etal_2018}, and recent disk observations \citep{Carrascogonzalez_etal_2019, Tazzari_etal_2021, Zagaria_etal_2025}. We also remind the reader that the simulations performed by \citet{Li_Youdin_2021}, \citet{Lim_etal_2025b, Lim_etal_2025a}, and those presented here are axisymmetric, and that their clustering characteristics may differ from those derived from fully 3D simulations. 
\REV{As discussed in Sec.~\ref{RDFs}, this dimensionality is reflected in the kinetic energy spectra and can be accounted for in constructing the RDF comparison. We have taken the appropriate steps to do so here, although future work should continue to examine this effect carefully.}

\section{Summary Conclusions and Final Remarks}


\NewEditv{In this study we have investigated the {\textit{axisymmetric}} weakly turbulent state of settled particle layers near the disk midplane of a \REV{globally laminar} protoplanetary disk in a local shearing-box setup, focusing on the characterization of the properties of the self-generated turbulence within this settling particle layer under conditions where the classical large-scale azimuthal filaments arising from the SI either do not form \REV{at all}, or the simulation is analyzed at an early stage before such structures have had time to develop \citep[][]{Yang_etal_2017,Li_Youdin_2021, Lim_etal_2025a}. 

Nonetheless, \REV{as discussed in Sec. \ref{TCalways},} we observe the robust and widespread emergence of small-scale particle clumping consistent with the predictions of turbulent concentration (TC). \REV{Specifically,} we have observed short filamentary structures bounding extended gas-only voids that exhibit coherent vortical patterning. The particles are expelled from these regions of coherent vorticity and congregate along regions of high gas strain-rate.



After testing particles with different $\St_K$, 
we find the 
effective Stokes numbers \REV{scaled to} 
the local vorticity within void regions consistently center around $\St_\omega \sim 0.4-0.7$. \REV{These $\St_\omega$ values are notably near the \REV{significant value} of $\St_\ell = 0.3$}, where $\St_\ell$ denotes the effective turbulent Stokes number on scale $\ell$ inferred from kinetic energy spectral analysis—{\it i.e.}, the largest scale $\ell$ at which particle intermittency is predicted to be largest, 
as identified in prior statistical studies of TC \citep[][]{Hartlep_etal_2017,Hartlep_Cuzzi_2020}. The apparent convergence of $\St_\omega$ \REV{into the vicinity of this value across all simulations} \REV{required a novel use of the actual, non-Kolmogorov shape of the kinetic energy spectrum (Sec. \ref{RDFs})}. 
Whether this behavior persists with similar clarity in fully 3D simulations remains to be seen.  

Our numerical experiments have demonstrated that 
whenever a particle effectively ``feels like" it has a Stokes number of $\order 1$ while being inside \REV{a gas-vortex driven particle void}, it most rapidly gets expelled from it, in line with the centrifuge mechanism originally proposed by \citet[][also commonly known as {\textit{Maxey centrifuge}}]{Maxey_1987}, \REV{and also discussed by} 
\citet{Squires_Eaton_1990,Squires_Eaton_1991, Bragg_Collins_2014}. 


We have further examined \REV{in Sec. \ref{sec:driving_mechanism}} the question of what drives the turbulence in the three simulations analyzed here. For the two simulations with $\St_K \ll 1$, where the midplane-averaged particle-to-gas mass ratios ($\epsilon$) fall below unity, upper-bound estimates for the SI growth rate ($\sigma_{{\mathrm{SI}}}$) are 1–2 orders of magnitude smaller than the corresponding large-eddy overturn frequencies ($\Omega_L$). Since $\Omega_L$ characterizes the rate of turbulent deformation at the injection scale, it is highly unlikely that the SI could be dynamically relevant -- its growth is simply too slow to compete with the turbulent stirring at the injection scales. By contrast, the SymI presents a more plausible driver: its characteristic growth rate $\sigma{_{\mathrm{SymI}}}$ is found to be comparable to $\Omega_L$. While this possibility has been previously proposed  in \citetalias{Sengupta_Umurhan_2023}, a definitive demonstration remains outstanding. Still, \REV{our} timescale \REV{and lengthscale} analysis 
lends strong support to this hypothesis.


Whatever the driver of turbulence in these kinds of scenarios, we find that TC is a persistent \REV{self-generated} feature of turbulence \REV{in settled} particle layers and remains active whenever such turbulence is present. It \REV{seems plausible} 
that the small-scale fluctuations exceeding Roche density -- frequently reported in simulations showing large-scale, low-number of nearly axisymmetric, high-density filaments attributed to the SI \citep[e.g.,][]{Li_Youdin_2021, Lim_etal_2025a} -- are 
expressions of TC. Notably, the \REV{largest} fluctuations occur within the larger-scale filaments, and we suggest their amplified strength reflects the elevated local particle density established by the filaments themselves.}

\vspace{0.2in}

\NewEditv{The authors thank Anders Johansen, and Heloise M\'eheut for their constructive feedback, corrections, and the addition of previously omitted references to earlier versions of this manuscript.}
O.M.U., D.S., T.H., and P.R.E. acknowledge support from NASA Emerging worlds grant \#80NSSC25K7022 entitled ``Turbulent Concentration and the First Planetesimals". D.S. also acknowledges support from NASA Emerging Worlds grant titled  ``Planetesimal Formation: A Natural Synergy Between Streaming Instability and Turbulent Concentration.'' \#80NSSC24K1282.  All the simulations presented in this paper are performed on the NASA Advanced Supercomputing  (NAS) facility with generous computational resources provided through EW allocations.



\appendix

\renewcommand{\thefigure}{A\arabic{figure}}
\setcounter{figure}{0}

\section{Typical velocity differences between particles and gas}
\label{sec:appendixUpg}
\REV{
We confront the issue laid out in the main text in two ways. First, we note that near the midplane the characteristic radial particle drift velocities are generally much larger than the corresponding mean radial gas velocities. To assess whether the void vortices can nevertheless influence particle motion, we compare the particle drift speeds with the characteristic velocities of the voids themselves.  We begin by calculating the radially averaged particle velocity, $\overline u_{p}(z)$, as a function of height above the midplane. These profiles are shown in Figure \ref{fig:mean_radial_particle_velocities}. We then compare them with a characteristic void velocity, denoted $U_{\rm void}$. The voids typically span length scales of order $\sim 0.0025H$ to $\sim 0.005H$, with smaller voids generally occurring for smaller $\St_K$, corresponding roughly to wavenumbers in the range $k \sim 1200/H$ to $2000/H$. We estimate $U_{\rm void}$ from the gas kinetic energy spectrum shown in Fig. \ref{fig_U2_Rprime} by reading off the spectral velocities associated with these scales. In this way, we adopt the following conservative estimates: $U^2_{{\rm void}}(\St_K = 0.01) \approx 5\times 10^{-6} c_s^2$,
$U^2_{{\rm void}}(\St_K = 0.02) \approx 10\times 10^{-6} c_s^2$, and 
$U^2_{{\rm void}}(\St_K = 0.04) \approx 25\times 10^{-6} c_s^2$.  
\par
Inspection of Fig. \ref{fig:mean_radial_particle_velocities} shows that the corresponding particle-void gas velocities exceed the particle drift velocities by a factor of roughly $2$--$3$. Thus, particles residing in or near voids experience local gas motions that are at least comparable to, and generally larger than, their mean radial drift speeds. 
Under these circumstances, particles with effective Stokes numbers $\St_\omega=\order 1$ would be expected to undergo substantial deflection during the time they spend traversing the coherent vortical gas flow contained within a particle void. 
Indeed, this is one of the hallmark features of TC itself: particles with $\St_\omega=\order 1$ respond most strongly to coherent vortical structures and are preferentially expelled from these vortex cores, leading to concentration in the surrounding regions.
\par
The second approach we take is to assess directly how well coupled the particles are to the turbulent gas motions. Assuming that the turbulence injection scale, $L$, and corresponding large-eddy frequency, $\Omega_L$, are known, there exists an analytic framework—originally developed by \citet{Volk_etal_1980} and later refined and tested against numerical turbulence experiments by \citet{Cuzzi_Hogan_2003}—for estimating the typical velocity difference between particles obeying Epstein drag and the surrounding turbulent gas.

To quantify this coupling, we define
\beq
U_{pg}/U_{{\rm rms}} \equiv
\overline{\left|\mathbf{U'}-\mathbf{V'}\right|}/U_{{\rm rms}},
\eeq
where $U_{{\rm rms}}$ is the root-mean-square gas velocity fluctuation (e.g., Fig. \ref{fig_U2_Rprime}), while $\mathbf{U'}$ and $\mathbf{V'}$ are the fluctuating gas and particle velocity vectors, respectively. The overline denotes an average over the particle-bearing region of the domain. Smaller values of $U_{pg}/U_{{\rm rms}}$ indicate stronger particle-gas coupling.

The expression estimating $U_{pg}/U_{{\rm rms}}$ reported by \citet{Cuzzi_Hogan_2003} assumes a Kolmogorov inertial-range spectrum with index $n=-5/3$. However, the spectra measured in our simulations are generally steeper, with $n<-2$. In this regime, the integral appearing in the general formulation does not admit a simple analytic expression. In Appendix \ref{sec:appendixTurb} we therefore revisit the derivation of \citet{Cuzzi_Hogan_2003} and develop an asymptotic formulation valid for $n\leq -5/3$ in the limit $\St_L\ll1$. The general expression is given in Eq. (\ref{general_Upg_analytic}), while in the limit $\St_L<1$ and $n\le -5/3$ it reduces to
\beq
U^2_{pg}/U^2_{{\rm rms}} \approx
-\frac{\St_L^2}{4+2n} + 
\pi\frac{5+3n}{(3+n)^2}
\csc\left(\frac{4\pi}{3+n}\right)\St_L^{-2\frac{1+n}{3+n}}.
\label{Upg}
\eeq
For $\St_L\ll1$ and $n<-2$, the first term on the equation's righhand side asymptotically dominates the second one.

We compare this prediction directly against measurements from the simulations. To do so, we define the spatially dependent quantity
\beq
\Xi\left(\setF\right)\equiv
\sqrt{
\frac{(u_g-u_p)^2}{\overline u_g^2}
+\frac{(v_g-v_p)^2}{\overline v_g^2}
+\frac{(w_g-w_p)^2}{\overline w_g^2}
},
\eeq
evaluated only on the set of grid points containing particles, $\setF$. Each simulation therefore yields a distribution of values of $\Xi(\setF)$. If the analytic framework above accurately captures particle-gas coupling in these flows, then the peak of this distribution, which we denote by $\Xi_{\rm peak}$, should provide an estimate of $U_{pg}/U_{\rm rms}$ and thus be directly comparable with either the asymptotic prediction of Eq. (\ref{Upg}) or the exact expression given by Eq. (\ref{general_Upg_analytic}).
\begin{figure*}[ht]
\begin{center}
\includegraphics[width=0.75\textwidth]{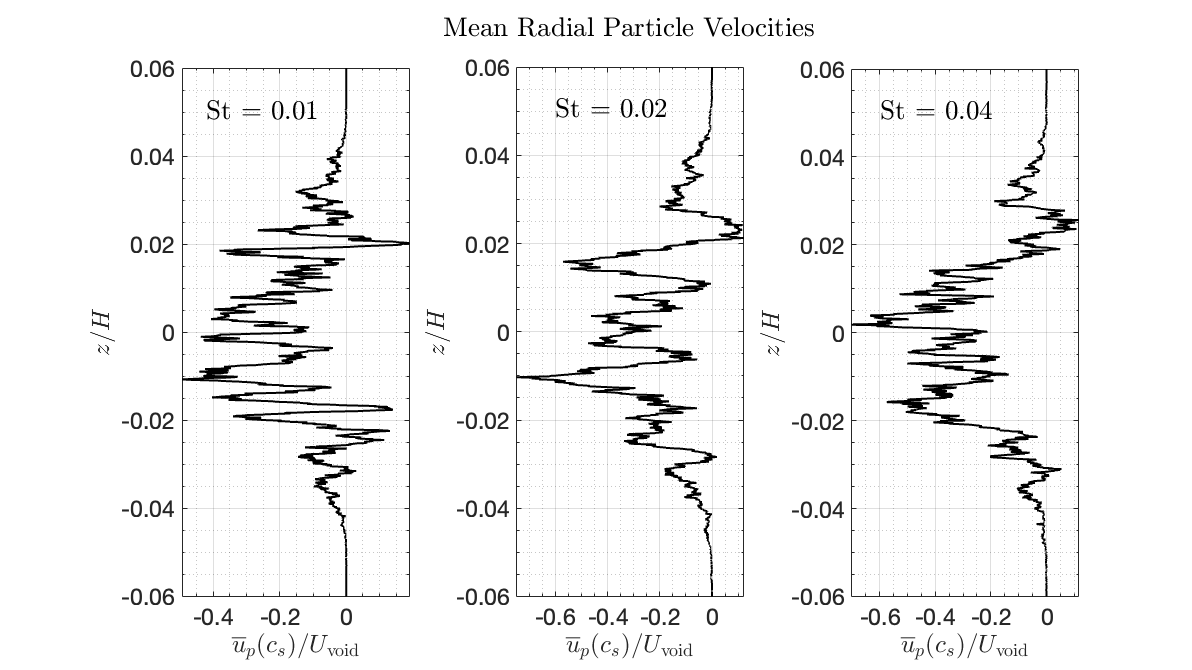}
\par
\end{center}
\vspace{-0.2in}
\caption{\REV{Mean radial particle velocities scaled by the corresponding gas vorticity speeds within particle voids for all three simulations.}}
\label{fig:mean_radial_particle_velocities}
\end{figure*}

\begin{figure}[ht]
\begin{center}
\includegraphics[width=0.5\textwidth]{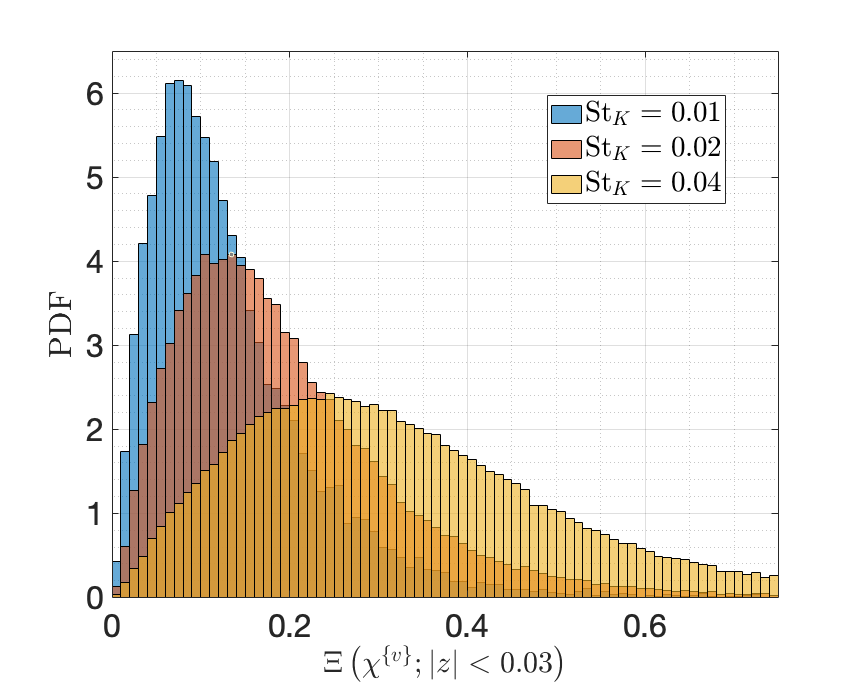}
\par
\end{center}
\vspace{-0.2in}
\caption{\REV{The PDF for $\Xi(\setF)$ for the three simulations performed.  In all histograms we have restricted our attention to the midplane region, i.e., $|z|<0.03$.  The peak values of the distribution are roughly: $\Xi_{{\rm peak}}(\St_K = 0.01) \approx 0.07$,
$\Xi_{{\rm peak}}(\St_K = 0.02) \approx 0.11$,
$\Xi_{{\rm peak}}(\St_K = 0.04) \approx 0.24$.  These compare favorably and are consistent with predicted trends for $U_{pg}/U_{{\rm rms}}$ (see  Table \ref{tbl:Upg_Xi_peak}).}
}
\label{fig:Xi_pdf}
\end{figure}

In Figure \ref{fig:Xi_pdf} we display the resulting distributions found in our simulations.  We find that the particles are in fact well coupled to the turbulent gas motions in which the values of $\Xi_{{\rm peak}}$
are significantly less than 1 and trend toward smaller values as $\St_L$ gets smaller.  These values also compare favorably with the predictions for $U_{pg}/U_{{\rm rms}}$ based on Eq. (\ref{Upg}), as demonstrated in Table \ref{tbl:Upg_Xi_peak}.  

Because $U_{pg}/U_{\rm rms}$ attains values of only $\order{0.1}$--$\order{0.2}$, the particles remain strongly coupled to the turbulent gas motions. We therefore consider it unlikely that particles with $\St_\omega=\order{1}$ pass through the coherent vortical structures with sufficient rapidity to avoid substantial deflection by them.

\begin{deluxetable}{c|cccc} 
\label{tbl:Upg_Xi_peak}
\tablecaption{Comparing $U_{pg}$ and $\Xi_{{\rm peak}}$ }
\tablehead{\colhead{$\St_K$} & \colhead{$\St_L$$^a$}  
&  \colhead{$n$$^b$}
&\colhead{$\Xi_{{\rm peak}}$} 
&
\colhead{$U_{pg}/U_{{\rm rms}}$$^c$}
}
\startdata 
0.01 & 0.03 & -2.0 & 0.07 &   0.07 \\
0.02 & 0.05 & -2.1 & 0.11 &   0.09 \\
0.04 & 0.155 & -2.1 & 0.24 &   0.20 \\
\hline
\enddata 
\vspace{0.05in}
 $^{a}$\! From Table \ref{tbl:simulation},
 $^b$\! Read from Fig. \ref{fig_All_Spectra},
$^c$\! Using Eq. (\ref{general_Upg_analytic}).
\end{deluxetable}
}

\setcounter{equation}{0}
\section{General Turbulent Particle Velocities}
\label{sec:appendixTurb}
\REV{
Extending prior analyses like those of \citet{Volk_etal_1978,Volk_etal_1980}, \citet{Cuzzi_Hogan_2003} formulated a framework for turbulent particle–gas velocities in nebular conditions with Epstein-drag coupling.  The framework is built upon calculating velocity autocorrelation functions to develop $U^2, U_p^2$ and $U_{pg}^2$, i.e., mean velocity fluctuations for the gas, particles, and the typical difference fluctuations between particles and gas, all respectively.  It is true that 
$U_{pg}^2 = U^2 - U_p^2$, and that a reasonable expression is
\beq
U_{pg}^2 = 2\int_{k_L}^{\infty}\varepsilon(k)\left(
\frac{1}{1 + t_\ell/t_s}\right)^{m+1} dk,
\eeq
\citep[][see their Eq. 9]{Cuzzi_Hogan_2003},
where $t_\ell = 1/\Omega_\ell$ is the eddy overturn time of eddy with size $\ell = 1/k$ (as discussed in the text in Section 3.3 and Eqs. \ref{eqn:stokesell}-\ref{eqn:omegal}).  The authors reported that $m=1$ best fits autocorrelation statistics from 3-D gas--particle turbulence (their Fig.~1). Because their focus was 3-D turbulence—with $\varepsilon = \epsilon_0k^{n}$ and $n=-5/3$—their results were quoted for that value. Our axisymmetric runs yield $n<-5/3$, so we re-evaluate the integral with $n$ left general. As such, the above can be rewritten in these general terms, where the integral is expressed terms of the variable $\chi \equiv k/k_L$, and making use of the definitions $U^2 \equiv \int_{k_L}^{\infty}2\varepsilon(k)dk$
and $\St_L \equiv \sqrt{2\epsilon_0 k_L^3}t_s$,
\beq
\displaystyle
U_{pg}^2 = U^2
\int_1^\infty \chi^n \left(\frac{\St_L}{\St_L + \chi^{-\frac{n+3}{2}}}\right)^2 d\chi,
\eeq
which, with the help of Mathematica, exactly integrates to the general result
\beq
U_{pg}^2\big/U^2 = \frac{2\St_L}{(3+n)(1+\St_L)}
+ 2\frac{(5+3n)}{(3+n)^2} 
\left(-\frac{1}{\St_L}\right)^{-\frac{4}{3+n}}
\frac{1}{\St_L^2}\beta\left(-1/\St_L; -2+\frac{4}{n+3},0\right),
\label{general_Upg_analytic}
\eeq
where $\beta(x; a,b)$ is the incomplete beta function.  We Taylor Series expand the above expression in a power series for small $\St_L$ and keep the first three terms to find
for $n\le-5/3$
\beq
U_{pg}^2\big/U^2 \approx
-\frac{\St_L^2}{4+2n} + \frac{2\St_L^3}{3+n}  + \pi\frac{5+3n}{(3+n)^2} \csc\left(\frac{4\pi}{3+n}\right)\St_L^{-2\frac{1+n}{3+n}},
\label{approximate_Upg_analytic}
\eeq
keeping in mind that evaluating the above expressions around $n=-2$ and $n=-5/3$ requires further Taylor series expansions.\!\footnote{For example, the leading term behavior around $n=-2$ is $\propto \St_L^2\ln\St_L$.}
We note that when $n>-2$ the second term asymptotically dominates the first since the exponent on the $\St_L$ term is always less than 2.}

\bibliography{reference}{}

@ARTICLE{Lin_2021,
       author = {{Lin}, Min-Kai},
        title = "{Stratified and Vertically Shearing Streaming Instabilities in Protoplanetary Disks}",
      journal = {\apj},
         year = 2021,
        month = feb,
       volume = {907},
       number = {2},
          eid = {64},
        pages = {64},
          doi = {10.3847/1538-4357/abcd9b},
archivePrefix = {arXiv},
       eprint = {2011.12300},
 primaryClass = {astro-ph.EP},
       adsurl = {https://ui.adsabs.harvard.edu/abs/2021ApJ...907...64L}
}

@ARTICLE{Ishitsu_etal_2009,
       author = {{Ishitsu}, Naoki and {Inutsuka}, Shu-ichiro and {Sekiya}, Minoru},
        title = "{Two-fluid Instability of Dust and Gas in the Dust Layer of a Protoplanetary Disk}",
      journal = {arXiv e-prints},
         year = 2009,
        month = may,
          eid = {arXiv:0905.4404},
        pages = {arXiv:0905.4404},
          doi = {10.48550/arXiv.0905.4404},
archivePrefix = {arXiv},
       eprint = {0905.4404},
 primaryClass = {astro-ph.EP},
       adsurl = {https://ui.adsabs.harvard.edu/abs/2009arXiv0905.4404I}
}

@ARTICLE{Cuzzi_Hogan_2003,
       author = {{Cuzzi}, Jeffrey N. and {Hogan}, Robert C.},
        title = "{Blowing in the wind. I. Velocities of chondrule-sized particles in a turbulent protoplanetary nebula}",
      journal = {\icarus},
         year = 2003,
        month = jul,
       volume = {164},
       number = {1},
        pages = {127-138},
          doi = {10.1016/S0019-1035(03)00104-0},
       adsurl = {https://ui.adsabs.harvard.edu/abs/2003Icar..164..127C}
}

@ARTICLE{Balachandar_Eaton_2010,
       author = {{Balachandar}, S. and {Eaton}, John K.},
        title = "{Turbulent Dispersed Multiphase Flow}",
      journal = {Annual Review of Fluid Mechanics},
         year = 2010,
        month = jan,
       volume = {42},
        pages = {111-133},
          doi = {10.1146/annurev.fluid.010908.165243},
       adsurl = {https://ui.adsabs.harvard.edu/abs/2010AnRFM..42..111B}
}

@ARTICLE{Baronett_etal_2024,
       author = {{Baronett}, Stanley A. and {Yang}, Chao-Chin and {Zhu}, Zhaohuan},
        title = "{Dust-gas dynamics driven by the streaming instability with various pressure gradients}",
      journal = {\mnras},
         year = 2024,
        month = mar,
       volume = {529},
       number = {1},
        pages = {275-295},
          doi = {10.1093/mnras/stae272},
archivePrefix = {arXiv},
       eprint = {2401.10430},
 primaryClass = {astro-ph.EP},
       adsurl = {https://ui.adsabs.harvard.edu/abs/2024MNRAS.529..275B}
}

@ARTICLE{Bec_etal_2010,
       author = {{Bec}, J. and {Biferale}, L. and {Lanotte}, A.~S. and {Scagliarini}, A. and {Toschi}, F.},
        title = "{Turbulent pair dispersion of inertial particles}",
      journal = {Journal of Fluid Mechanics},
         year = 2010,
        month = feb,
       volume = {645},
        pages = {497},
          doi = {10.1017/S0022112009992783},
archivePrefix = {arXiv},
       eprint = {0904.2314},
 primaryClass = {physics.flu-dyn},
       adsurl = {https://ui.adsabs.harvard.edu/abs/2010JFM...645..497B}
}

@ARTICLE{Bec_etal_2023,
       author = {{Bec}, J. and {Gustavsson}, K. and {Mehlig}, B.},
        title = "{Statistical Models for the Dynamics of Heavy Particles in Turbulence}",
      journal = {Annual Review of Fluid Mechanics},
         year = 2023,
        month = oct,
       volume = {56},
        pages = {189-213},
          doi = {10.1146/annurev-fluid-032822-014140},
archivePrefix = {arXiv},
       eprint = {2304.01312},
 primaryClass = {physics.flu-dyn},
       adsurl = {https://ui.adsabs.harvard.edu/abs/2023AnRFM..56..189B}
}

@ARTICLE{Bragg_Collins_2014,
       author = {{Bragg}, Andrew D. and {Collins}, Lance R.},
        title = "{New insights from comparing statistical theories for inertial particles in turbulence: I. Spatial distribution of particles}",
      journal = {New Journal of Physics},
         year = 2014,
        month = may,
       volume = {16},
       number = {5},
          eid = {055013},
        pages = {055013},
          doi = {10.1088/1367-2630/16/5/055013},
       adsurl = {https://ui.adsabs.harvard.edu/abs/2014NJPh...16e5013B}
}

@ARTICLE{Carrascogonzalez_etal_2019,
       author = {{Carrasco-Gonz{\'a}lez}, Carlos and {Sierra}, Anibal and {Flock}, Mario and {Zhu}, Zhaohuan and {Henning}, Thomas and {Chandler}, Claire and {Galv{\'a}n-Madrid}, Roberto and {Mac{\'\i}as}, Enrique and {Anglada}, Guillem and {Linz}, Hendrik and {Osorio}, Mayra and {Rodr{\'\i}guez}, Luis F. and {Testi}, Leonardo and {Torrelles}, Jos{\'e} M. and {P{\'e}rez}, Laura and {Liu}, Yao},
        title = "{The Radial Distribution of Dust Particles in the HL Tau Disk from ALMA and VLA Observations}",
      journal = {\apj},
         year = 2019,
        month = sep,
       volume = {883},
       number = {1},
          eid = {71},
        pages = {71},
          doi = {10.3847/1538-4357/ab3d33},
archivePrefix = {arXiv},
       eprint = {1908.07140},
 primaryClass = {astro-ph.EP},
       adsurl = {https://ui.adsabs.harvard.edu/abs/2019ApJ...883...71C}
}

@ARTICLE{Carrera_etal_2015,
       author = {{Carrera}, Daniel and {Johansen}, Anders and {Davies}, Melvyn B.},
        title = "{How to form planetesimals from mm-sized chondrules and chondrule aggregates}",
      journal = {\aap},
         year = 2015,
        month = jul,
       volume = {579},
          eid = {A43},
        pages = {A43},
          doi = {10.1051/0004-6361/201425120},
archivePrefix = {arXiv},
       eprint = {1501.05314},
 primaryClass = {astro-ph.EP},
       adsurl = {https://ui.adsabs.harvard.edu/abs/2015A&A...579A..43C}
}

@ARTICLE{Chen_Lin_2020,
       author = {{Chen}, Kan and {Lin}, Min-Kai},
        title = "{How Efficient Is the Streaming Instability in Viscous Protoplanetary Disks?}",
      journal = {\apj},
         year = 2020,
        month = mar,
       volume = {891},
       number = {2},
          eid = {132},
        pages = {132},
          doi = {10.3847/1538-4357/ab76ca},
archivePrefix = {arXiv},
       eprint = {2002.07188},
 primaryClass = {astro-ph.EP},
       adsurl = {https://ui.adsabs.harvard.edu/abs/2020ApJ...891..132C}
}

@ARTICLE{Li_Youdin_2021,
       author = {{Li}, Rixin and {Youdin}, Andrew N.},
        title = "{Thresholds for Particle Clumping by the Streaming Instability}",
      journal = {\apj},
         year = 2021,
        month = oct,
       volume = {919},
       number = {2},
          eid = {107},
        pages = {107},
          doi = {10.3847/1538-4357/ac0e9f},
archivePrefix = {arXiv},
       eprint = {2105.06042},
 primaryClass = {astro-ph.EP},
       adsurl = {https://ui.adsabs.harvard.edu/abs/2021ApJ...919..107L}
}

@ARTICLE{Squire_Hopkins_2020,
       author = {{Squire}, Jonathan and {Hopkins}, Philip F.},
        title = "{Physical models of streaming instabilities in protoplanetary discs}",
      journal = {\mnras},
         year = 2020,
        month = oct,
       volume = {498},
       number = {1},
        pages = {1239-1251},
          doi = {10.1093/mnras/staa2311},
archivePrefix = {arXiv},
       eprint = {2003.01738},
 primaryClass = {astro-ph.EP},
       adsurl = {https://ui.adsabs.harvard.edu/abs/2020MNRAS.498.1239S}
}

@ARTICLE{Chiang_2008,
       author = {{Chiang}, E.},
        title = "{Vertical Shearing Instabilities in Radially Shearing Disks: The Dustiest Layers of the Protoplanetary Nebula}",
      journal = {\apj},
         year = 2008,
        month = mar,
       volume = {675},
       number = {2},
        pages = {1549-1558},
          doi = {10.1086/527354},
archivePrefix = {arXiv},
       eprint = {0711.4349},
 primaryClass = {astro-ph},
       adsurl = {https://ui.adsabs.harvard.edu/abs/2008ApJ...675.1549C}
}

@ARTICLE{Cuzzi_etal_1993,
       author = {{Cuzzi}, Jeffrey N. and {Dobrovolskis}, Anthony R. and {Champney}, Joelle M.},
        title = "{Particle-Gas Dynamics in the Midplane of a Protoplanetary Nebula}",
      journal = {\icarus},
         year = 1993,
        month = nov,
       volume = {106},
       number = {1},
        pages = {102-134},
          doi = {10.1006/icar.1993.1161},
       adsurl = {https://ui.adsabs.harvard.edu/abs/1993Icar..106..102C}
}

@ARTICLE{Cuzzi_etal_2001,
       author = {{Cuzzi}, Jeffrey N. and {Hogan}, Robert C. and {Paque}, Julie M. and {Dobrovolskis}, Anthony R.},
        title = "{Size-selective Concentration of Chondrules and Other Small Particles in Protoplanetary Nebula Turbulence}",
      journal = {\apj},
         year = 2001,
        month = jan,
       volume = {546},
       number = {1},
        pages = {496-508},
          doi = {10.1086/318233},
archivePrefix = {arXiv},
       eprint = {astro-ph/0009210},
 primaryClass = {astro-ph},
       adsurl = {https://ui.adsabs.harvard.edu/abs/2001ApJ...546..496C}
}

@ARTICLE{Cuzzi_etal_2003,
       author = {{Cuzzi}, Jeffrey N. and {Davis}, Sanford S. and
         {Dobrovolskis}, Anthony R.},
        title = "{Blowing in the wind. II. Creation and redistribution of refractory inclusions in a turbulent protoplanetary nebula}",
      journal = {\icarus},
         year = "2003",
        month = "Dec",
       volume = {166},
       number = {2},
        pages = {385-402},
          doi = {10.1016/j.icarus.2003.08.016},
       adsurl = {https://ui.adsabs.harvard.edu/abs/2003Icar..166..385C}
}

@ARTICLE{Cuzzi_etal_2008,
       author = {{Cuzzi}, Jeffrey N. and {Hogan}, Robert C. and {Shariff}, Karim},
        title = "{Toward Planetesimals: Dense Chondrule Clumps in the Protoplanetary Nebula}",
      journal = {\apj},
         year = 2008,
        month = nov,
       volume = {687},
       number = {2},
        pages = {1432-1447},
          doi = {10.1086/591239},
archivePrefix = {arXiv},
       eprint = {0804.3526},
 primaryClass = {astro-ph},
       adsurl = {https://ui.adsabs.harvard.edu/abs/2008ApJ...687.1432C}
}

@ARTICLE{Cuzzi_etal_2010,
       author = {{Cuzzi}, Jeffrey N. and {Hogan}, Robert C. and {Bottke}, William F.},
        title = "{Towards initial mass functions for asteroids and Kuiper Belt Objects}",
      journal = {\icarus},
         year = "2010",
        month = "Aug",
       volume = {208},
       number = {2},
        pages = {518-538},
          doi = {10.1016/j.icarus.2010.03.005},
archivePrefix = {arXiv},
       eprint = {1004.0270},
 primaryClass = {astro-ph.EP},
       adsurl = {https://ui.adsabs.harvard.edu/abs/2010Icar..208..518C}
}

@ARTICLE{Dobrovolskis_etal_1999,
       author = {{Dobrovolskis}, Anthony R. and {Dacles-Mariani}, Jennifer S. and {Cuzzi}, Jeffrey N.},
        title = "{Production and damping of turbulence by particles in the solar nebula}",
      journal = {\jgr},
         year = 1999,
        month = jan,
       volume = {104},
       number = {E12},
        pages = {30805-30816},
          doi = {10.1029/1999JE001053},
       adsurl = {https://ui.adsabs.harvard.edu/abs/1999JGR...10430805D}
}

@ARTICLE{Dubrulle_etal_1995,
       author = {{Dubrulle}, B. and {Morfill}, G. and {Sterzik}, M.},
        title = "{The dust subdisk in the protoplanetary nebula.}",
      journal = {\icarus},
         year = 1995,
        month = apr,
       volume = {114},
       number = {2},
        pages = {237-246},
          doi = {10.1006/icar.1995.1058},
       adsurl = {https://ui.adsabs.harvard.edu/abs/1995Icar..114..237D}
}

@ARTICLE{Elghobasi_Truesdell_1992,
       author = {{Elghobashi}, S. and {Truesdell}, G.~C.},
        title = "{Direct simulation of particle dispersion in a decaying isotropic turbulence}",
      journal = {Journal of Fluid Mechanics},
         year = 1992,
        month = sep,
       volume = {242},
        pages = {655-700},
          doi = {10.1017/S0022112092002532},
       adsurl = {https://ui.adsabs.harvard.edu/abs/1992JFM...242..655E}
}

@ARTICLE{Estrada_etal_2016,
       author = {{Estrada}, Paul R. and {Cuzzi}, Jeffrey N. and {Morgan}, Demitri A.},
        title = "{Global Modeling of Nebulae with Particle Growth, Drift, and Evaporation Fronts. I. Methodology and Typical Results}",
      journal = {\apj},
         year = "2016",
        month = "Feb",
       volume = {818},
       number = {2},
          eid = {200},
        pages = {200},
          doi = {10.3847/0004-637X/818/2/200},
archivePrefix = {arXiv},
       eprint = {1506.01420},
 primaryClass = {astro-ph.EP},
       adsurl = {https://ui.adsabs.harvard.edu/abs/2016ApJ...818..200E}
}

@ARTICLE{Estrada_etal_2022,
       author = {{Estrada}, Paul R. and {Cuzzi}, Jeffrey N. and {Umurhan}, Orkan M.},
        title = "{Global Modeling of Nebulae with Particle Growth, Drift, and Evaporation Fronts. II. The Influence of Porosity on Solids Evolution}",
      journal = {\apj},
         year = 2022,
        month = sep,
       volume = {936},
       number = {1},
          eid = {42},
        pages = {42},
          doi = {10.3847/1538-4357/ac7ffd},
archivePrefix = {arXiv},
       eprint = {2207.12626},
 primaryClass = {astro-ph.EP},
       adsurl = {https://ui.adsabs.harvard.edu/abs/2022ApJ...936...42E}
}

@ARTICLE{Hartlep_etal_2017,
       author = {{Hartlep}, Thomas and {Cuzzi}, Jeffrey N. and {Weston}, Brian},
        title = "{Scale dependence of multiplier distributions for particle concentration, enstrophy, and dissipation in the inertial range of homogeneous turbulence}",
      journal = {\pre},
         year = "2017",
        month = "Mar",
       volume = {95},
       number = {3},
          eid = {033115},
        pages = {033115},
          doi = {10.1103/PhysRevE.95.033115},
archivePrefix = {arXiv},
       eprint = {1703.05871},
 primaryClass = {physics.flu-dyn},
       adsurl = {https://ui.adsabs.harvard.edu/abs/2017PhRvE..95c3115H}
}

@INPROCEEDINGS{Hartlep_Cuzzi_2017,
       author = {{Hartlep}, T. and {Cuzzi}, J.~N.},
        title = "{A Path to Chondrule Aggregates}",
    booktitle = {Accretion: Building New Worlds Conference},
         year = 2017,
       editor = {{LPI Editorial Board}},
       series = {LPI Contributions},
       volume = {2043},
        month = aug,
          eid = {2049},
        pages = {2049},
       adsurl = {https://ui.adsabs.harvard.edu/abs/2017LPICo2043.2049H}
}

@ARTICLE{Hartlep_Cuzzi_2020,
       author = {{Hartlep}, Thomas and {Cuzzi}, Jeffrey N.},
        title = "{Cascade Model for Planetesimal Formation by Turbulent Clustering}",
      journal = {\apj},
         year = 2020,
        month = apr,
       volume = {892},
       number = {2},
          eid = {120},
        pages = {120},
          doi = {10.3847/1538-4357/ab76c3},
archivePrefix = {arXiv},
       eprint = {2002.06321},
 primaryClass = {astro-ph.EP},
       adsurl = {https://ui.adsabs.harvard.edu/abs/2020ApJ...892..120H}
}

@ARTICLE{Hopkins_Lee_2016,
       author = {{Hopkins}, Philip F. and {Lee}, Hyunseok},
        title = "{The fundamentally different dynamics of dust and gas in molecular clouds}",
      journal = {\mnras},
         year = 2016,
        month = mar,
       volume = {456},
       number = {4},
        pages = {4174-4190},
          doi = {10.1093/mnras/stv2745},
archivePrefix = {arXiv},
       eprint = {1510.02477},
 primaryClass = {astro-ph.GA},
       adsurl = {https://ui.adsabs.harvard.edu/abs/2016MNRAS.456.4174H}
}

@ARTICLE{Johansen_etal_2007,
       author = {{Johansen}, Anders and {Oishi}, Jeffrey S. and {Mac Low}, Mordecai-Mark and
         {Klahr}, Hubert and {Henning}, Thomas and {Youdin}, Andrew},
        title = "{Rapid planetesimal formation in turbulent circumstellar disks}",
      journal = {\nat},
         year = "2007",
        month = "Aug",
       volume = {448},
       number = {7157},
        pages = {1022-1025},
          doi = {10.1038/nature06086},
archivePrefix = {arXiv},
       eprint = {0708.3890},
 primaryClass = {astro-ph},
       adsurl = {https://ui.adsabs.harvard.edu/abs/2007Natur.448.1022J}
}

@ARTICLE{Kolmogorov_1941,
       author = {{Kolmogorov}, A.},
        title = "{The Local Structure of Turbulence in Incompressible Viscous Fluid for Very Large Reynolds' Numbers}",
      journal = {Akademiia Nauk SSSR Doklady},
         year = 1941,
        month = jan,
       volume = {30},
        pages = {301-305},
       adsurl = {https://ui.adsabs.harvard.edu/abs/1941DoSSR..30..301K}
}

@ARTICLE{Lim_etal_2025b,
       author = {{Lim}, Jeonghoon and {Baronett}, Stanley A. and {Simon}, Jacob B. and {Yang}, Chao-Chin and {Sengupta}, Debanjan and {Umurhan}, Orkan M. and {Lyra}, Wladimir},
        title = "{Bridging Unstratified and Stratified Simulations of the Streaming Instability for $τ_s=0.1$ Grains}",
      journal = {arXiv e-prints},
         year = 2025,
        month = may,
          eid = {arXiv:2505.23902},
        pages = {arXiv:2505.23902},
          doi = {10.48550/arXiv.2505.23902},
archivePrefix = {arXiv},
       eprint = {2505.23902},
 primaryClass = {astro-ph.EP},
       adsurl = {https://ui.adsabs.harvard.edu/abs/2025arXiv250523902L}
}

@ARTICLE{Lim_etal_2025a,
       author = {{Lim}, Jeonghoon and {Simon}, Jacob B. and {Li}, Rixin and {Carrera}, Daniel and {Baronett}, Stanley A. and {Youdin}, Andrew N. and {Lyra}, Wladimir and {Yang}, Chao-Chin},
        title = "{Probing Conditions for Strong Clumping by the Streaming Instability: Small Dust Grains and Low Dust-to-gas Density Ratio}",
      journal = {\apj},
         year = 2025,
        month = mar,
       volume = {981},
       number = {2},
          eid = {160},
        pages = {160},
          doi = {10.3847/1538-4357/adb311},
archivePrefix = {arXiv},
       eprint = {2410.17319},
 primaryClass = {astro-ph.EP},
       adsurl = {https://ui.adsabs.harvard.edu/abs/2025ApJ...981..160L}
}

@ARTICLE{Volk_etal_1980,
       author = {{V\"olk}, H.~J. and {Jones}, F.~C. and {Morfill}, G.~E. and {Roeser}, S.},
        title = "{Collisions between Grains in a Turbulent Gas}",
      journal = {\aap},
         year = 1980,
        month = may,
       volume = {85},
       number = {3},
        pages = {316-325},
       adsurl = {https://ui.adsabs.harvard.edu/abs/1980A&A....85..316V}
}

@ARTICLE{Volk_etal_1978,
       author = {{V\"olk}, H.~J. and {Morfill}, G. and {Roeser}, S. and {Jones}, F.~C.},
        title = "{Induced velocities of grains embedded in a turbulent gas}",
      journal = {Moon and Planets},
         year = 1978,
        month = oct,
       volume = {19},
       number = {2},
        pages = {221-227},
          doi = {10.1007/BF00896995},
       adsurl = {https://ui.adsabs.harvard.edu/abs/1978M&P....19..221V}
}

@article{Maxey_1987,
        title={The gravitational settling of aerosol particles in homogeneous turbulence and random flow fields},
        volume={174},
        DOI={10.1017/S0022112087000193},
        journal={Journal of Fluid Mechanics},
        author={Maxey, M. R.},
        year={1987},
        pages={441–465}
}

@ARTICLE{Pan_etal_2011,
       author = {{Pan}, Liubin and {Padoan}, Paolo and {Scalo}, John and {Kritsuk}, Alexei G. and {Norman}, Michael L.},
        title = "{Turbulent Clustering of Protoplanetary Dust and Planetesimal Formation}",
      journal = {\apj},
         year = 2011,
        month = oct,
       volume = {740},
       number = {1},
          eid = {6},
        pages = {6},
          doi = {10.1088/0004-637X/740/1/6},
archivePrefix = {arXiv},
       eprint = {1106.3695},
 primaryClass = {astro-ph.EP},
       adsurl = {https://ui.adsabs.harvard.edu/abs/2011ApJ...740....6P}
}

@ARTICLE{Pencil_code,
       author = {{Pencil Code Collaboration} and {Brandenburg}, Axel and {Johansen}, Anders and {Bourdin}, Philippe and {Dobler}, Wolfgang and {Lyra}, Wladimir and {Rheinhardt}, Matthias and {Bingert}, Sven and {Haugen}, Nils and {Mee}, Antony and {Gent}, Frederick and {Babkovskaia}, Natalia and {Yang}, Chao-Chin and {Heinemann}, Tobias and {Dintrans}, Boris and {Mitra}, Dhrubaditya and {Candelaresi}, Simon and {Warnecke}, J{\"o}rn and {K{\"a}pyl{\"a}}, Petri and {Schreiber}, Andreas and {Chatterjee}, Piyali and {K{\"a}pyl{\"a}}, Maarit and {Li}, Xiang-Yu and {Kr{\"u}ger}, Jonas and {Aarnes}, J{\o}rgen and {Sarson}, Graeme and {Oishi}, Jeffrey and {Schober}, Jennifer and {Plasson}, Rapha{\"e}l and {Sandin}, Christer and {Karchniwy}, Ewa and {Rodrigues}, Luiz and {Hubbard}, Alexander and {Guerrero}, Gustavo and {Snodin}, Andrew and {Losada}, Illa and {Pekkil{\"a}}, Johannes and {Qian}, Chengeng},
        title = "{The Pencil Code, a modular MPI code for partial differential equations and particles: multipurpose and multiuser-maintained}",
      journal = {The Journal of Open Source Software},
         year = 2021,
        month = feb,
       volume = {6},
       number = {58},
          eid = {2807},
        pages = {2807},
          doi = {10.21105/joss.02807},
archivePrefix = {arXiv},
       eprint = {2009.08231},
 primaryClass = {astro-ph.IM},
       adsurl = {https://ui.adsabs.harvard.edu/abs/2021JOSS....6.2807P}
}

@ARTICLE{Sekiya_1998,
       author = {{Sekiya}, Minoru},
        title = "{Quasi-Equilibrium Density Distributions of Small Dust Aggregations in the Solar Nebula}",
      journal = {\icarus},
         year = 1998,
        month = jun,
       volume = {133},
       number = {2},
        pages = {298-309},
          doi = {10.1006/icar.1998.5933},
       adsurl = {https://ui.adsabs.harvard.edu/abs/1998Icar..133..298S}
}

@ARTICLE{Sengupta_etal_2019,
       author = {{Sengupta}, Debanjan and {Dodson-Robinson}, Sarah E. and
         {Hasegawa}, Yasuhiro and {Turner}, Neal J.},
        title = "{Growth and Settling of Dust Particles in Protoplanetary Nebulae: Implications for Opacity, Thermal Profile, and Gravitational Instability}",
      journal = {\apj},
         year = 2019,
        month = mar,
       volume = {874},
       number = {1},
          eid = {26},
        pages = {26},
          doi = {10.3847/1538-4357/aafc36},
archivePrefix = {arXiv},
       eprint = {1808.03016},
 primaryClass = {astro-ph.EP},
       adsurl = {https://ui.adsabs.harvard.edu/abs/2019ApJ...874...26S}
}

@ARTICLE{Sengupta_Umurhan_2023,
       author = {{Sengupta}, Debanjan and {Umurhan}, Orkan M.},
        title = "{Turbulence in Particle-laden Midplane Layers of Planet-forming Disks}",
      journal = {\apj},
         year = 2023,
        month = jan,
       volume = {942},
       number = {2},
          eid = {74},
        pages = {74},
          doi = {10.3847/1538-4357/ac9411},
archivePrefix = {arXiv},
       eprint = {2209.11205},
 primaryClass = {astro-ph.EP},
       adsurl = {https://ui.adsabs.harvard.edu/abs/2023ApJ...942...74S}
}

@ARTICLE{Sengupta_etal_2024,
       author = {{Sengupta}, Debanjan and {Cuzzi}, Jeffrey N. and {Umurhan}, Orkan M. and {Lyra}, Wladimir},
        title = "{Length and Velocity Scales in Protoplanetary Disk Turbulence}",
      journal = {\apj},
         year = 2024,
        month = may,
       volume = {966},
       number = {1},
          eid = {90},
        pages = {90},
          doi = {10.3847/1538-4357/ad2c89},
archivePrefix = {arXiv},
       eprint = {2402.15475},
 primaryClass = {astro-ph.EP},
       adsurl = {https://ui.adsabs.harvard.edu/abs/2024ApJ...966...90S}
}

@ARTICLE{Simon_etal_2018,
       author = {{Simon}, J.~I. and {Cuzzi}, J.~N. and {McCain}, K.~A. and {Cato}, M.~J. and
         {Christoffersen}, P.~A. and {Fisher}, K.~R. and {Srinivasan}, P. and
         {Tait}, A.~W. and {Olson}, D.~M. and {Scargle}, J.~D.},
        title = "{Particle size distributions in chondritic meteorites: Evidence for pre-planetesimal histories}",
      journal = {Earth and Planetary Science Letters},
         year = "2018",
        month = "Jul",
       volume = {494},
        pages = {69-82},
          doi = {10.1016/j.epsl.2018.04.021},
       adsurl = {https://ui.adsabs.harvard.edu/abs/2018E&PSL.494...69S}
}

@ARTICLE{Stamper_Taylor_2017,
       author = {{Stamper}, Megan A. and {Taylor}, John R.},
        title = "{The transition from symmetric to baroclinic instability in the Eady model}",
      journal = {Ocean Dynamics},
         year = 2017,
        month = jan,
       volume = {67},
       number = {1},
        pages = {65-80},
          doi = {10.1007/s10236-016-1011-6},
       adsurl = {https://ui.adsabs.harvard.edu/abs/2017OcDyn..67...65S}
}

@ARTICLE{Squires_Eaton_1990,
       author = {{Squires}, Kyle D. and {Eaton}, John K.},
        title = "{Particle response and turbulence modification in isotropic turbulence}",
      journal = {Physics of Fluids A},
         year = 1990,
        month = jul,
       volume = {2},
       number = {7},
        pages = {1191-1203},
          doi = {10.1063/1.857620},
       adsurl = {https://ui.adsabs.harvard.edu/abs/1990PhFlA...2.1191S}
}

@ARTICLE{Squires_Eaton_1991,
       author = {{Squires}, Kyle D. and {Eaton}, John K.},
        title = "{Preferential concentration of particles by turbulence}",
      journal = {Physics of Fluids A},
         year = 1991,
        month = may,
       volume = {3},
       number = {5},
        pages = {1169-1178},
          doi = {10.1063/1.858045},
       adsurl = {https://ui.adsabs.harvard.edu/abs/1991PhFlA...3.1169S}
}

@ARTICLE{Hopkins_2016b,
       author = {{Hopkins}, Philip F.},
        title = "{Jumping the gap: the formation conditions and mass function of `pebble-pile' planetesimals}",
      journal = {\mnras},
         year = 2016,
        month = mar,
       volume = {456},
       number = {3},
        pages = {2383-2405},
          doi = {10.1093/mnras/stv2820},
archivePrefix = {arXiv},
       eprint = {1401.2458},
 primaryClass = {astro-ph.EP},
       adsurl = {https://ui.adsabs.harvard.edu/abs/2016MNRAS.456.2383H}
}

@ARTICLE{Gerosa_etal_2023,
       author = {{Gerosa}, Fabiola A. and {M{\'e}heut}, H{\'e}lo{\"\i}se and {Bec}, J{\'e}r{\'e}mie},
        title = "{Clusters of heavy particles in two-dimensional Keplerian turbulence}",
      journal = {European Physical Journal Plus},
         year = 2023,
        month = jan,
       volume = {138},
       number = {1},
          eid = {9},
        pages = {9},
          doi = {10.1140/epjp/s13360-022-03585-8},
archivePrefix = {arXiv},
       eprint = {2210.13147},
 primaryClass = {astro-ph.EP},
       adsurl = {https://ui.adsabs.harvard.edu/abs/2023EPJP..138....9G}
}

@ARTICLE{Hopkins_2016a,
       author = {{Hopkins}, Philip F.},
        title = "{A simple phenomenological model for grain clustering in turbulence}",
      journal = {\mnras},
         year = 2016,
        month = jan,
       volume = {455},
       number = {1},
        pages = {89-111},
          doi = {10.1093/mnras/stv2226},
archivePrefix = {arXiv},
       eprint = {1307.7147},
 primaryClass = {astro-ph.EP},
       adsurl = {https://ui.adsabs.harvard.edu/abs/2016MNRAS.455...89H}
}

@ARTICLE{Squire_Hopkins_2018a,
       author = {{Squire}, Jonathan and {Hopkins}, Philip F.},
        title = "{Resonant drag instabilities in protoplanetary discs: the streaming instability and new, faster growing instabilities}",
      journal = {\mnras},
         year = 2018,
        month = jul,
       volume = {477},
       number = {4},
        pages = {5011-5040},
          doi = {10.1093/mnras/sty854},
archivePrefix = {arXiv},
       eprint = {1711.03975},
 primaryClass = {astro-ph.EP},
       adsurl = {https://ui.adsabs.harvard.edu/abs/2018MNRAS.477.5011S}
}

@ARTICLE{Tazzari_etal_2021,
       author = {{Tazzari}, M. and {Clarke}, C.~J. and {Testi}, L. and {Williams}, J.~P. and {Facchini}, S. and {Manara}, C.~F. and {Natta}, A. and {Rosotti}, G.},
        title = "{Multiwavelength continuum sizes of protoplanetary discs: scaling relations and implications for grain growth and radial drift}",
      journal = {\mnras},
         year = 2021,
        month = sep,
       volume = {506},
       number = {2},
        pages = {2804-2823},
          doi = {10.1093/mnras/stab1808},
archivePrefix = {arXiv},
       eprint = {2010.02249},
 primaryClass = {astro-ph.EP},
       adsurl = {https://ui.adsabs.harvard.edu/abs/2021MNRAS.506.2804T}
}

@ARTICLE{Umurhan_etal_2020,
 	 author = {Umurhan, O. M. and Estrada, P. R. and Cuzzi, J. N.},
        title = "{Streaming Instability in Turbulent Protoplanetary Disks}",
      journal = {\apj},
         year = "2020",
        month = "May",
       volume = {895},
       number = {1},
        pages = {26},
        doi = {10.3847/1538-4357/ab899d}
}

@ARTICLE{Wang_Squires_1996,
       author = {{Wang}, Qunzhen and {Squires}, Kyle D.},
        title = "{Large eddy simulation of particle-laden turbulent channel flow}",
      journal = {Physics of Fluids},
         year = 1996,
        month = may,
       volume = {8},
       number = {5},
        pages = {1207-1223},
          doi = {10.1063/1.868911},
       adsurl = {https://ui.adsabs.harvard.edu/abs/1996PhFl....8.1207W}
}

@ARTICLE{Weidenschilling_1980,
       author = {{Weidenschilling}, S.~J.},
        title = "{Dust to planetesimals: Settling and coagulation in the solar nebula}",
      journal = {\icarus},
         year = 1980,
        month = oct,
       volume = {44},
       number = {1},
        pages = {172-189},
          doi = {10.1016/0019-1035(80)90064-0},
       adsurl = {https://ui.adsabs.harvard.edu/abs/1980Icar...44..172W}
}

@ARTICLE{Yang_Lei_1998,
       author = {{Yang}, C.~Y. and {Lei}, U.},
        title = "{The role of the turbulent scales in the settling velocity of heavy particles in homogeneous isotropic turbulence}",
      journal = {Journal of Fluid Mechanics},
         year = 1998,
        month = sep,
       volume = {371},
       number = {1},
        pages = {179-205},
          doi = {10.1017/S0022112098002328},
       adsurl = {https://ui.adsabs.harvard.edu/abs/1998JFM...371..179Y}
}

@ARTICLE{Yang_etal_2017,
       author = {{Yang}, C. -C. and {Johansen}, A. and {Carrera}, D.},
        title = "{Concentrating small particles in protoplanetary disks through the streaming instability}",
      journal = {\aap},
         year = "2017",
        month = "Oct",
       volume = {606},
          eid = {A80},
        pages = {A80},
          doi = {10.1051/0004-6361/201630106},
archivePrefix = {arXiv},
       eprint = {1611.07014},
 primaryClass = {astro-ph.EP},
       adsurl = {https://ui.adsabs.harvard.edu/abs/2017A&A...606A..80Y}
}

@ARTICLE{Youdin_Goodman_2005,
       author = {{Youdin}, Andrew N. and {Goodman}, Jeremy},
        title = "{Streaming Instabilities in Protoplanetary Disks}",
      journal = {\apj},
         year = 2005,
        month = feb,
       volume = {620},
       number = {1},
        pages = {459-469},
          doi = {10.1086/426895},
archivePrefix = {arXiv},
       eprint = {astro-ph/0409263},
 primaryClass = {astro-ph},
       adsurl = {https://ui.adsabs.harvard.edu/abs/2005ApJ...620..459Y}
}

@ARTICLE{Youdin_Johansen_2007,
       author = {{Youdin}, A. and {Johansen}, A.},
        title = "{Protoplanetary Disk Turbulence Driven by the Streaming Instability: Linear Evolution and Numerical Methods}",
      journal = {\apj},
         year = 2007,
        month = jun,
       volume = {662},
       number = {1},
        pages = {613-626},
          doi = {10.1086/516729},
archivePrefix = {arXiv},
       eprint = {astro-ph/0702625},
 primaryClass = {astro-ph},
       adsurl = {https://ui.adsabs.harvard.edu/abs/2007ApJ...662..613Y}
}

@ARTICLE{Youdin_Lithwick_2007,
       author = {{Youdin}, Andrew N. and {Lithwick}, Yoram},
        title = "{Particle stirring in turbulent gas disks: Including orbital oscillations}",
      journal = {\icarus},
         year = 2007,
        month = dec,
       volume = {192},
       number = {2},
        pages = {588-604},
          doi = {10.1016/j.icarus.2007.07.012},
archivePrefix = {arXiv},
       eprint = {0707.2975},
 primaryClass = {astro-ph},
       adsurl = {https://ui.adsabs.harvard.edu/abs/2007Icar..192..588Y}
}

@ARTICLE{Zagaria_etal_2025,
       author = {{Zagaria}, Francesco and {Facchini}, Stefano and {Curone}, Pietro and {Williams}, Jonathan P. and {Clarke}, Cathie J. and {Ribas}, {\'A}lvaro and {Tazzari}, Marco and {Mac{\'\i}as}, Enrique and {Booth}, Richard A. and {Rosotti}, Giovanni P. and {Testi}, Leonardo},
        title = "{Multi-frequency analysis of the ALMA and VLA high resolution continuum observations of the substructured disc around CI Tau. Preference for sub-mm-sized low-porosity amorphous carbonaceous grains}",
      journal = {arXiv e-prints},
         year = 2025,
        month = jul,
          eid = {arXiv:2507.08797},
        pages = {arXiv:2507.08797},
          doi = {10.48550/arXiv.2507.08797},
archivePrefix = {arXiv},
       eprint = {2507.08797},
 primaryClass = {astro-ph.EP},
       adsurl = {https://ui.adsabs.harvard.edu/abs/2025arXiv250708797Z}
}

@ARTICLE{Zaichik_Alipchenkov_2009,
       author = {{Zaichik}, Leonid I. and {Alipchenkov}, Vladimir M.},
        title = "{Statistical models for predicting pair dispersion and particle clustering in isotropic turbulence and their applications}",
      journal = {New Journal of Physics},
         year = 2009,
        month = oct,
       volume = {11},
       number = {10},
          eid = {103018},
        pages = {103018},
          doi = {10.1088/1367-2630/11/10/103018},
       adsurl = {https://ui.adsabs.harvard.edu/abs/2009NJPh...11j3018Z}
}
\bibliographystyle{aasjournal}
\end{document}